\documentclass[11pt]{article}
\usepackage{graphicx}
\usepackage{dcolumn}
\usepackage{color}
\newcommand{\BABARPubYear}    {02}

\newcommand{\BABARConfNumber} {023}
\newcommand{\SLACPubNumber} {9323}

\input pubboard/babarsym

\def\Kmaybestar {\ensuremath{K^{(*)}\xspace}}

\def\kll {\B\to\Kmaybestar\ellell\xspace}

\def\modekee {\ensuremath{B^\pm\rightarrow K^\pm\epem}}

\def\modekll {\ensuremath{B^\pm\rightarrow K^\pm\ellell}}
\def\modekmm {\ensuremath{B^\pm\rightarrow K^\pm\mumu}}

\def\modeksee {\ensuremath{B^0\rightarrow K^0_{\scriptscriptstyle S}\epem}}

\def\modeksll {\ensuremath{B^0\rightarrow K^0_{\scriptscriptstyle S}\ellell}}
\def\modeksmm {\ensuremath{B^0\rightarrow K^0_{\scriptscriptstyle S}\mumu}}

\def\modekstavgll {\ensuremath{B\rightarrow K^{*}\ellell}}

\def\modekstkee {\ensuremath{B^0\rightarrow K^{*0}\epem}}

\def\modekstkll {\ensuremath{B^0\rightarrow K^{*0}\ellell}}
\def\modekstkmm {\ensuremath{B^0\rightarrow K^{*0}\mumu}}

\def\modekstksee {\ensuremath{B^\pm\rightarrow K^{*\pm}\epem}}

\def\modekstksll {\ensuremath{B^\pm\rightarrow K^{*\pm}\ellell}}
\def\modekstksmm {\ensuremath{B^\pm\rightarrow K^{*\pm}\mumu}}

\newlength{\digitspace}
\newlength{\hundredspace}
\newlength{\tenspace}
\long\def\inst#1{\par\nobreak\kern 4pt\nobreak
    {\it #1}\par\vskip 10pt plus 3pt minus 3pt}

\begin{document}
{\pagestyle{empty}

\begin{flushright}
\babar-CONF-\BABARPubYear/\BABARConfNumber \\
SLAC-PUB-\SLACPubNumber \\
July 2002 \\
\end{flushright}

\par\vskip 5cm

\begin{center}
\Large \bf \boldmath
Evidence for the Flavor Changing Neutral Current Decays 
$B \rightarrow K \ell^+ \ell^-$ and  $ B \rightarrow K^{\ast} \ell^+ \ell^-$
\end{center}
\bigskip

\begin{center}
\large The \babar\ Collaboration\\
\mbox{ }\\
July 24, 2002
\end{center}
\bigskip \bigskip

\begin{center}
\large \bf Abstract
\end{center}

We present preliminary results from a search for the rare, flavor-changing
neutral current decays
$B\to K\ell^+\ell^-$ and $B\to K^*\ell^+\ell^-$,
where $\ell^+\ell^-$ is either an $e^+e^-$ or $\mu^+\mu^-$ pair.
The data sample comprises $(84.4\pm0.9)\times 10^6$
$\Upsilon(4S)\to \BB$ decays (77.8 fb$^{-1}$) collected with the 
\babar\ detector at the \pep2 \BF. 
For $B\to K\ell^+\ell^-$, we observe a signal with estimated significance of
$4.4\sigma$ 
and obtain ${\mathcal B}(B\to K\ell^+\ell^-) =
(0.78^{+0.24+0.11}_{-0.20-0.18})\times 10^{-6}$ (averaged over 
$\ell = e$ and $\mu$).  For $B\to K^*\ell^+\ell^-$,
we observe an excess of events over background with estimated significance of
$2.8\sigma$.
We obtain ${\mathcal B}(B\to K^*\ell^+\ell^-) = 
(1.68^{+0.68}_{-0.58}\pm0.28)\times 10^{-6}$ and the 90\% C.L.~upper limit
${\mathcal B}(B\to K^*\ell^+\ell^-)<3.0\times 10^{-6}$.

\vfill

\begin{center}
Contributed to the 31$^{st}$ International Conference on High Energy Physics,\\ 
7/24---7/31/2002, Amsterdam, The Netherlands
\end{center}

\vspace{1.0cm}
\begin{center}
{\em Stanford Linear Accelerator Center, Stanford University, 
Stanford, CA 94309} \\ \vspace{0.1cm}\hrule\vspace{0.1cm}
Work supported in part by Department of Energy contract DE-AC03-76SF00515.
\end{center}

\newpage
} 

\begin{center}
\small

The \babar\ Collaboration,
\bigskip

B.~Aubert,
D.~Boutigny,
J.-M.~Gaillard,
A.~Hicheur,
Y.~Karyotakis,
J.~P.~Lees,
P.~Robbe,
V.~Tisserand,
A.~Zghiche
\inst{Laboratoire de Physique des Particules, F-74941 Annecy-le-Vieux, France }
A.~Palano,
A.~Pompili
\inst{Universit\`a di Bari, Dipartimento di Fisica and INFN, I-70126 Bari, Italy }
% ...

\end{center}\newpage

\section{Introduction}
\label{sec:Introduction}

The flavor-changing neutral current decays 
$B\to K\ell^+\ell^-$ and $B\to K^*(892)\ell^+\ell^-$, 
where $\ell^{\pm}$ is 
a charged lepton, are highly suppressed in the Standard 
Model, with branching fractions 
predicted
to be of order 
$10^{-7}-10^{-6}$~\cite{bib:TheoryA,bib:TheoryB,bib:AliUpdate}. The dominant contributions arise at
one-loop level and include both electroweak penguin and box diagrams.
Besides probing Standard Model loop effects, these rare decays are important
because their rates and kinematic distributions are sensitive
to new, heavy particles---such as those predicted by 
supersymmetric models---that can appear virtually in the 
loops~\cite{bib:TheoryA,bib:TheoryB,bib:AliUpdate}. 

Searches for $B\to K^{(*)}\ell^+\ell^-$ decays have been performed by
\babar~\cite{bib:BaBarKll}, Belle~\cite{bib:Belle}, CDF~\cite{bib:CDF}, and CLEO~\cite{bib:CLEO}. 
Using a sample of 29 fb$^{-1}$, the Belle Collaboration
has observed a signal for the decay $B\to K\ell^+\ell^-$ with branching
fraction ${\cal B}(B\to K\ell^+\ell^-)=(0.75^{+0.25}_{-0.21}\pm0.09)\times 10^{-6}$,
averaged over electron and muon modes. The recently published \babar~\cite{bib:BaBarKll}
result was based on 
a 20.7 fb$^{-1}$ sample and yielded the 90\% C.L.~upper limits 
${\cal B}(B\to K\ell^+\ell^-)<0.51\times 10^{-6}$ and 
${\cal B}(B\to K^*\ell^+\ell^-)<3.1\times 10^{-6}$.
In the present study, we update our result with a total sample that
is nearly four times larger than that used in the original \babar\
analysis.

We search for the following decays:
$B^+\to K^+\ell^+\ell^-$, 
$B^0\to K_S^0\ell^+\ell^-$,
$B^+\to K^{*+}\ell^+\ell^-$, and 
$B^0\to K^{*0}\ell^+\ell^-$, where
$K^{*0}\to K^+\pi^-$, $K^{*+}\to K_S^0\pi^+$,
$K_S^0\to\pi^+\pi^-$, and
$\ell$ is either an 
$e$ or $\mu$.
Throughout this paper, charge-conjugate modes are implied.

\section{Theoretical predictions}
\label{sec:Theory}

\begin{figure}[b!]
\begin{center}
\includegraphics[height=5cm]{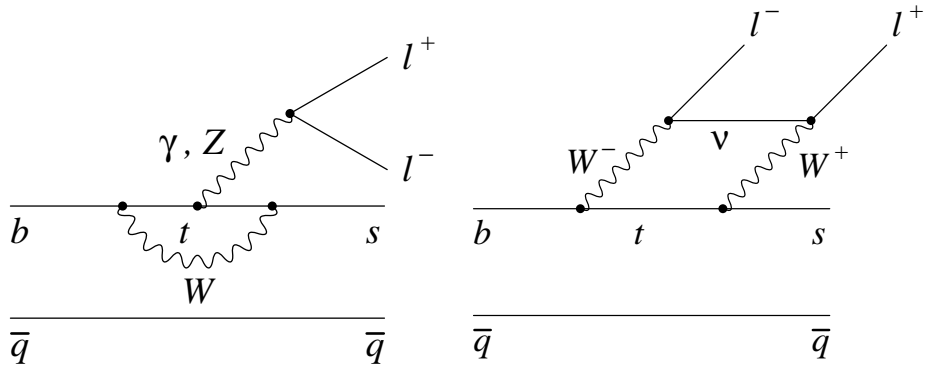}
\caption{Standard Model diagrams
for the decays $B\to K^{(*)}\ell^+\ell^-$.}
\label{fig:PenguinDiagrams}
\end{center}
\end{figure}

The Standard Model processes for $B\to K^{(*)}\ell^+\ell^-$
include three contributions: the electromagnetic (EM) and $Z$ penguin diagrams,
and the $W^+W^-$ box diagrams (Fig.~\ref{fig:PenguinDiagrams}). 
Evidence for the EM penguin amplitude was first obtained
by the CLEO experiment from the observation of the exclusive
decay $B\to K^*\gamma$
and later from the inclusive process $B\to X_s\gamma$, 
where $X_s$ is any hadronic system with 
strangeness~\cite{bib:CLEOKstargam,bib:CLEOXsgam}. 
Since then, several experiments have observed
these decays, 
and the Particle Data Group~\cite{bib:PDG2002} calculates the world averages
${\cal B}(B^0\to K^{*0}\gamma)=(4.3\pm0.4)\times 10^{-5}$ and  
${\cal B}(B\to X_s\gamma)=(3.3\pm0.4)\times 10^{-4}$.

The Standard Model rates for $B\to K^{*}\ell^+\ell^-$ 
and $B\to X_s\ell^+\ell^-$ 
are expected to be 25--80 times
smaller than those for $B\to K^*\gamma$ and $B\to X_s\gamma$.
For example, Ali {\it et al.}~predict~\cite{bib:AliUpdate} 
${\cal B}(B\to X_s e^+ e^-)=(6.89\pm1.01)\times 10^{-6}$ and
${\cal B}(B\to X_s\mu^+\mu^-)=(4.15\pm0.70)\times 10^{-6}$ 
for the inclusive branching fractions.
Exclusive final states 
have larger theoretical uncertainties but are experimentally
more accessible.
Ali {\it et al.}~\cite{bib:AliUpdate} predict
${\cal B}(B\to K\ell^+\ell^-)=(0.35\pm0.12)\times 10^{-6}$ for both $e^+e^-$ and $\mu^+\mu^-$
final states, 
${\cal B}(B\to K^* e^+e^-)=(1.58\pm0.49)\times 10^{-6}$, and 
${\cal B}(B\to K^* \mu^+\mu^-)=(1.19\pm0.39)\times 10^{-6}$. 

\begin{figure}[tb!]
\begin{center}
\includegraphics[height=12cm]{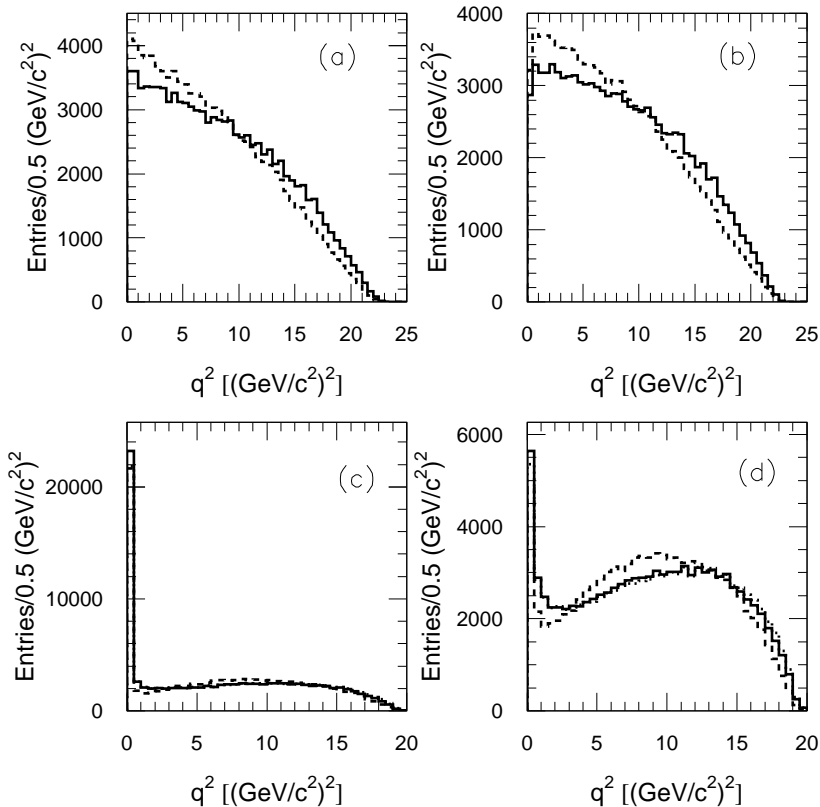}
\caption{Model predictions~\protect\cite{bib:TheoryA} for the decay rate as a function
of $q^2$ for (a) $B\to Ke^+e^-$ (b) $B\to K\mu^+\mu^-$
(c) $B\to K^* e^+e^-$ and (d) $B\to K^*\mu^+\mu^-$. The solid histogram
is based on the light cone QCD sum rules model of Ali {\it et al.}, the
dashed histogram on the QCD sum rules model of Colangelo {\it et al.},
and the dot-dashed histogram on the quark model of Melikhov, Nikitin,
and Simula. The distributions are normalized to the same area, so that
any differences in the overall rates are not shown.  
}
\label{fig:model_dep}
\end{center}
\end{figure}

Figure~\ref{fig:model_dep} shows the predicted rates as a function of $q^2\equiv m_{\ell^+\ell^-}^2$ 
for a number of theoretical models. The $B\to K\ell^+\ell^-$ modes
display a falloff in the rate from low values of $q^2$ (fast recoil kaon)
to high values (slow recoil kaon). (Although $B\to K\gamma$ is forbidden
by angular momentum conservation, $B\to K\ell^+\ell^-$ is allowed since $q^2> 0$.)
The $q^2$ distributions for $B\to K^*\ell^+\ell^-$ behave quite
differently, with large enhancements at low $q^2$, but otherwise
with the maximum rate at some intermediate $q^2$ value. 
The low $q^2$ enhancements are due to the EM penguin amplitude,
which has a pole at $q^2=0$, strongly
enhancing $B\to K^* e^+ e^-$ and, to a lesser extent,
$B\to K^*\mu^+\mu^-$. As a consequence, most Standard-Model-based predictions give a higher rate for 
$B\to K^* e^+e^-$ than  
for $B\to K^* \mu^+\mu^-$. 
At values of $q^2$ above this pole region,
the contributions from the $Z$ penguin and the $W^+W^-$ box diagram 
become very important. 

While the $B\to K^{(*)}\ell^+\ell^-$ decays are rarer than the
pure EM penguin processes, they
have sensitivity to three Wilson coefficients
$C_7$, $C_9$, and $C_{10}$ in the Operator
Product Expansion~\cite{bib:TheoryA}, whereas the $b\to s\gamma$ 
rate is 
sensitive mainly to the magnitude of a single Wilson coefficient, $C_7$.
New physics could, for example, change the sign of $C_7$
in $B\to K^*\ell^+\ell^-$; the resulting modification of the
interference terms with the other amplitudes could enhance the
rate by a factor of two over the Standard Model prediction.

Measurements of the kinematic distributions in
$B\to K^*\ell^+\ell^-$ decays will eventually provide a
probe of new physics with less theoretical uncertainty
than the decay rates.
Certain features
of the lepton angular distribution
as a function of $q^2$ can
be predicted reliably and can be dramatically modified by new physics.
In particular, the lepton forward-backward asymmetry in the
dilepton rest frame is predicted to become zero at a particular
value of $q^2$ in the Standard Model, but the sign of the asymmetry
and the location of the zero can be modified by new physics.
The study of these angular distributions will
require very large data samples and will be investigated by future 
experiments, including those at the LHC.

\section{The \babar\ detector and data sample}
\label{sec:babar}

The data used in the analysis were collected with the \babar\ detector
at the \pep2\ asymmetric energy $e^+e^-$ 
storage ring at the Stanford Linear Accelerator Center.
We analyze the data taken in the 1999--2002 runs, consisting
of a 77.8 \invfb\ sample accumulated on the $\Upsilon(4S)$ resonance, as well as 
9.6 \invfb\ taken at
a center-of-mass energy 40 MeV below the $\Upsilon(4S)$ resonance peak 
to obtain a pure continuum sample. 
Continuum events include non-resonant
$e^+e^-\to q\overline q$ production, where $q=u$, $d$, $s$, or $c$, as well
as two-photon and $\tau^+\tau^-$ events.                 
The on-resonance sample contains
$(84.4\pm 0.9)\times 10^6$ $\Upsilon(4S)\to \BB$ events.

The \babar\ detector is described in detail elsewhere~\cite{bib:babarNIM}. 
Of particular importance for this
analysis are the charged-particle tracking system and the detectors used for
particle identification. At radii between about 3 cm and 14 cm,
charged tracks are measured 
in a five-layer silicon vertex tracker (SVT).
Tracking beyond
the SVT is provided by the drift chamber, which extends in radius 
from 23.6 to 80.9 cm and provides up to 40 track-measurement points.
Just outside the drift chamber is the DIRC, 
a Cherenkov ring-imaging particle identification system. The DIRC
provides charged particle velocity information through the
measurement of the Cherenkov light cone opening angle.
Cherenkov light
is produced by charged particles as they pass through an array of 144
five-meter-long fused silica quartz bars. The Cherenkov light is transmitted to
the instrumented end of the bars by total internal reflection, preserving the
information on the angle of the light emission with respect to the
track direction. The DIRC is used for kaon identification in this
analysis and is essential to our background rejection.
Electrons are identified using
an electromagnetic calorimeter comprising 6580 thallium-doped CsI
crystals. These systems are mounted inside a 1.5~T solenoidal
superconducting magnet to provide momentum measurement in the
charged particle tracking systems.
Muons are identified using the Instrumented Flux Return (IFR),
in which resistive plate chambers (RPCs) are interleaved
with the iron plates of the magnet flux return.
 
\section{Analysis overview}
\label{sec:AnalysisOverview}

The most obvious analysis challenge is
that in each of the eight signal channels 
there are background processes with identical
final-state particles that proceed through 
charmonium resonances. For example, the decay
$B\to K^+\ell^+\ell^-$ has backgrounds from
$B\to J/\psi(\to\ell^+\ell^-)K^+$ and $B\to \psi(2S)(\to\ell^+\ell^-)K^+$
that must be carefully understood and vetoed. These
backgrounds are particularly difficult in
the electron channels, where bremsstrahlung in
detector material can easily cause the dilepton
mass to fall well below that of the mass of the charmonium
resonance. On the other hand, these backgrounds provide
nearly ideal control samples for each channel, allowing
us to make data vs.~Monte Carlo comparisons and checks of 
selection criteria efficiencies
in a way that carries over almost directly to the signal processes.
Before searching for a signal in any newly acquired data sample, we validate our
analysis by comprehensively studying the yields and shapes
of distributions in the charmonium control samples. Although
these processes have a narrow distribution in
$q^2$, they still allow us to confirm that the 
lepton identification efficiencies used in the Monte Carlo simulations 
are valid
over a broad range of momenta. We have therefore
studied the charmonium control samples in detail, not only
to design our vetoes, but also to establish signal shape
parameters and
to verify Monte Carlo predictions of signal
efficiencies.

The most important non-charmonium background contributions are from
$B\overline B$ backgrounds with either two real leptons 
or one real lepton and one hadron misidentified as a lepton (usually as a 
muon);
continuum processes, especially $c\overline c$ events with a pair of
$D\to K^{(*)}\ell^+\nu$ decays or events with hadrons faking leptons;
and small, but potentially very important contributions from 
a number of $B$ decay modes with similar
topology to the signal, such as $B^+\to D^0\pi^+$ with $D^0\to K^-\pi^+$,
in which hadrons misidentified as leptons
can create a false signal. Another example of a signal-like
background is $B\to K^*\gamma$ followed by
conversion of the photon into $e^+e^-$ in the detector material.
This process is distinct from the decay $B\to K^*e^+e^-$ at low $q^2$,
where the photon undergoes internal conversion. While no single
such background is large, it is important to establish
that the sum of many processes with very small contributions does
not produce a significant number of peaking background events.

In the $B\to K\ell^+\ell^-$ modes, the dominant background is
from continuum events. Much of this
background can be suppressed using event shape variables
that discriminate between the jet-like topology of continuum
events and the much more spherical topology of signal (and other 
$B\overline B$ events). In
$B\to K^*\ell^+\ell^-$ channels, the $B\overline B$ background is substantially 
larger than that from the continuum. 
Much of the $B\overline B$ background is due to pairs of 
primary semileptonic
decays from the $B$ and $\overline B$ mesons. Such events 
are characterized by large amounts of missing 
energy associated with the two neutrinos. We have constructed
two main analysis variables, one for $B\overline B$ rejection
($B$ likelihood)
and one for continuum-background rejection (Fisher
discriminant). Each of these
variables combines information from several
kinematic quantities.

To prevent bias in the analysis, we optimize the
event-selection criteria using (1) Monte Carlo samples
both signal and background, (2) the off-resonance
data, and (3) a ``large sideband'' region in the data that is
outside of the fit region (defined below). 
Signal efficiencies were
determined using the model of Ali {\it et al.}~\cite{bib:TheoryA},
although a set of other models~\cite{bib:TheoryA} was used to estimate
a systematic uncertainty on
the efficiencies due to model dependence. None of 
our models for the
signal efficiency 
include small effects arising from 
interference between the electroweak penguin
amplitudes and the amplitudes for the decays
involving charmonium intermediate
states. The interference is constructive on the 
lower side of the $J/\psi$ and $\psi(2S)$ charmonium resonance
masses but is destructive on
the upper side. The effect of this interference on our yields is
also suppressed by the
charmonium vetoes.

After event selection, we perform a two-dimensional unbinned
maximimum likelihood fit to the joint distribution in 
\begin{eqnarray}
m_{\rm ES}&=& \sqrt{E_{\rm b}^{*2} - (\sum_i {\bf p}^*_i)^2}\nonumber\\ 
\Delta E&=& \sum_i\sqrt{m_i^2 + {\bf p}_i^{*2}}- E_{\rm b}^*,\nonumber
\end{eqnarray}
where $E_{\rm b}^*$ is the beam energy in the $e^+e^-$ rest (CM) frame,
${\bf p}_i^*$ is the CM~momentum of daughter particle $i$ of the
$B$ meson candidate, and $m_i$ is the mass hypothesis for particle $i$. 
For signal events, $m_{\rm ES}$ peaks at the $B$ meson mass with 
a resolution of about 2.5 MeV$/c^2$ and $\Delta E$ peaks near zero,  
indicating that the candidate system of particles has total energy consistent with
the beam energy in the CM~frame.

\begin{figure}[!tb]
 \begin{center}
  \mbox{\hspace{0.3in}\includegraphics[width=\linewidth]{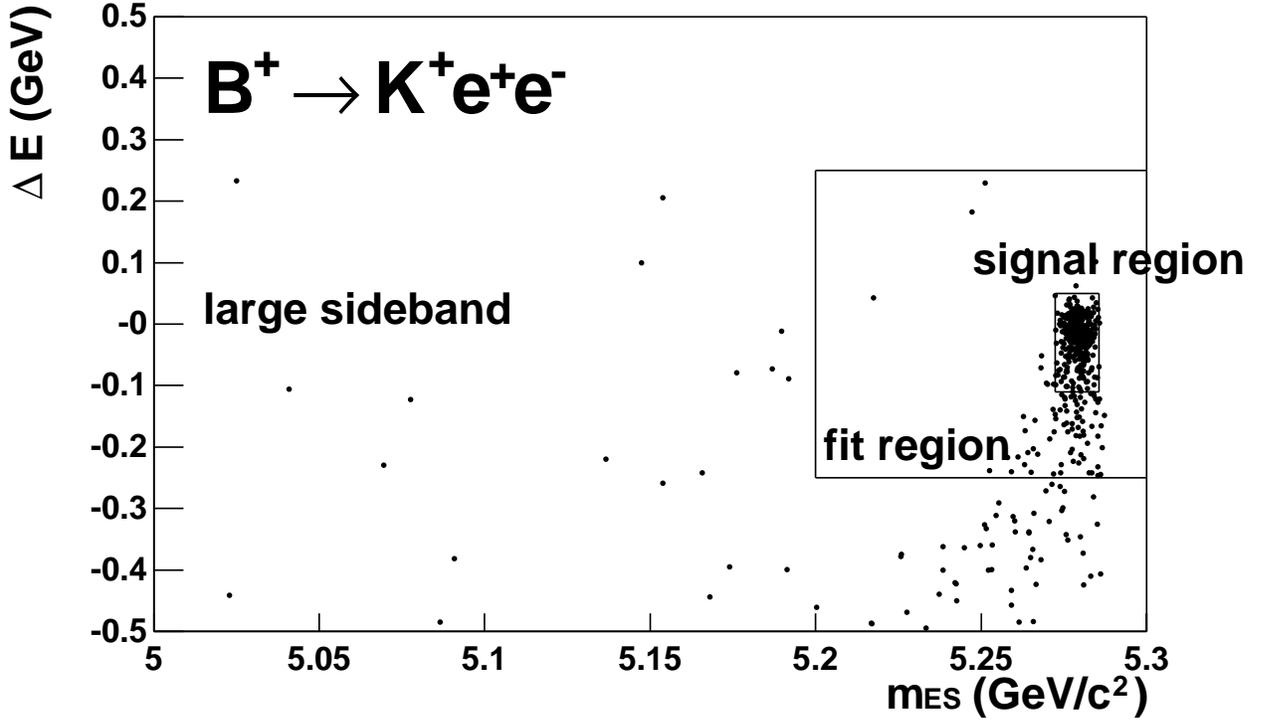}}
 \end{center}
 \vspace{-0.5cm}
\caption[Regions in the $\Delta E$ vs.~$m_{\rm ES}$ plane.]
{\label{fig:regions}
The distribution of Monte-Carlo-simulated $B^+\to K^+ e^+e^-$ 
events in the $m_{\rm ES}$ vs. $\Delta E$ plane. The nominal signal region, 
fit region, and large-sideband region are shown. The full fit
region in the data is concealed (``blinded'') until the event selection 
procedure is defined and the data validation steps are performed.
The nominal signal region is used only for the optimization of the
event selection criteria using Monte Carlo samples; the extraction of the 
signal yields from the data is based on an analysis
of the entire fit region.
}
\end{figure}

Figure~\ref{fig:regions} shows three regions that we have defined in the
$\Delta E$ vs.~$m_{\rm ES}$ plane. The events are taken from
a {\tt GEANT}4~\cite{bib:Geant4} simulation of the process $B^+\to K^+ e^+e^-.$ The small
box where most of the events lie is the nominal signal region. This
region is used to indicate where most of the signal is expected, but 
signal events are extracted over the full fit region, which is much
larger. The fit region is concealed during the process of determining
the event selection criteria.
The region outside the fit region is called the large sideband.
It is not concealed but is instead used to perform data vs.~Monte
Carlo comparisons in a region that is dominated by combinatorial
background. This region, together with samples of $Ke\mu$ and $K^*e\mu$
events, provides a way for us to study combinatorial backgrounds
without the possibility of biasing the sample of events in the 
fit region. In the fits, the background normalization is allowed
to float, as are other parameters describing the background shape.
Thus, while we use Monte Carlo samples to motivate our background
parametrizations, the parameter values themselves are not fixed from
the Monte Carlo simulation.

\section{Event selection}
\label{sec:EventSelection}

We select events that have at least four charged tracks,
the ratio $R_2$ of the second and zeroth Fox-Wolfram moments~\cite{bib:FoxWolfram} (calculated using charged tracks and unmatched calorimeter clusters)
less than 0.5, and 
two oppositely charged leptons with labratory-frame
momenta
\mbox{$p > 0.5 \ (1.0) \ {\rm GeV}/c$} for $e$ ($\mu$) 
candidates. The leptons are identified with strict criteria to 
minimize the misidentification rates for hadrons. 
The typical probability for a pion to be misidentified as an 
electron is 0.2\% to 0.3\%, depending on momentum; 
the probability for a pion to be misidentified
as a muon is 1.3\% to 2.7\%.  
Electron-positron pairs consistent with photon conversions
in the detector material are vetoed. In the $B\to K e^+ e^-$ modes,
$e^+e^-$ pairs at low $q^2$ are vetoed independent of the distance
of the vertex from the beam axis,
but in $B\to K^*e^+e^-$, where we expect substantial rate
at low $q^2$, we veto only conversion candidates that occur at a 
radius of at least 2 cm, where detector materials are present.
We require charged kaon candidates to be identified as kaons with
strict criteria and
the charged pion in $K^{*}\to K\pi$  not to be identified 
as a kaon. The use of strict criteria for kaon and lepton 
identification is necessary to suppress peaking backgrounds.
For \mbox{$B\rightarrow K^{\ast} \ell^+\ell^-$}, 
we require the mass of the $K^\ast$ candidate to 
be within \mbox{75$~\mevcc$} of the mean $K^\ast(892)$ mass.  
$K^0_S$ candidates are reconstructed from two oppositely charged tracks 
that form a good vertex
displaced from the primary vertex by at least 1 mm. The mass
of the $\pi^+\pi^-$ system must be within $\pm 9.3$ MeV/$c^2$ of the
nominal $K_S^0$ mass.

\begin{figure}[!tb]
 \begin{center}
   \includegraphics[width=\linewidth]{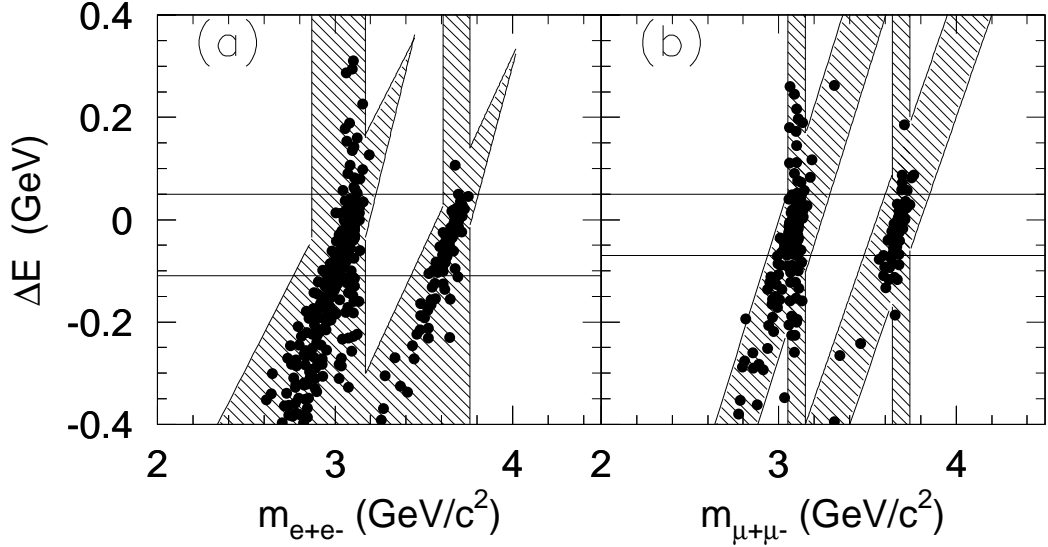}
  \end{center}
  \vspace{-0.5cm}
\caption[Definition of the charmonium veto region.]
{\label{fig:charmoniumveto}
Charmonium veto in the $\Delta E$ vs.~$m_{\ell^+\ell^-}$ plane
for (a) $B\to K^{(*)} e^+e^-$
and (b) $B\to K^{(*)} \mu^+\mu^-$. Hatched regions are vetoed.  
The dots correspond to a Monte Carlo simulation of 
$B\to J/\psi(\to\ell^+\ell^-)K$ and $B\to\psi(2S)(\to\ell^+\ell^-)K$.
Most signal events lie in the $\Delta E$ region
between the horizontal lines.
}
\end{figure}

The decays \mbox{$B \to J/\psi(\to \ell^+\ell^-)K^{(*)}$} and
\mbox{$B \to \psi(2S)(\to \ell^+\ell^-)K^{(*)}$} have identical 
topologies to signal events.
These backgrounds
are suppressed by applying a veto in 
the $\Delta E$ vs.~$m_{\ell^+\ell^-}$ plane (Fig.~\ref{fig:charmoniumveto}). 
This veto removes charmonium events not only with reconstructed 
$m_{\ell^+\ell^-}$ values 
near the nominal charmonium masses, but also events 
that lie
further away in $m_{\ell^+\ell^-}$ due to photon radiation (more 
pronounced in electron channels) 
or track mismeasurement. 
Removing these events---not only from the signal region, but also from
the full fit region---simplifies the description of the background shape.
Charmonium events can, however, escape this veto if one of the leptons (typically
a muon) and 
the kaon are misidentified as each other.
If reassignment of particle types results in a dilepton mass
consistent with the $J/\psi$ or $\psi(2S)$ mass, the candidate
is vetoed.
There is also significant feed-up from
\mbox{$B\to J/\psi K$} and \mbox{$B\to \psi(2S) K$} into
\mbox{$B \to K^\ast \ell^+ \ell^-$}, 
since energy lost due to bremsstrahlung in $B\to J/\psi K$ 
can be compensated for by including a random
pion. 
If the $K\ell^+\ell^-$ system in a $B\to K^*\ell^+\ell^-$ candidate is
kinematically consistent with $B\to J/\psi(\to\ell^+\ell^-\gamma) K$, assuming
that the photon (which is not directly observed)
was radiated along the direction of either lepton, then the 
candidate is vetoed. In modes with muons we also apply a veto against
backgrounds from $B\to D\pi$, with $D\to K^{(*)}\pi$. We reassign the 
particle hypothesis for the muon candidate with charge consistent
to form a $D$ when combined with the $K^{(*)}$ to a pion and veto
the candidate if the mass of the muon-$K^{(*)}$ system is
consistent with the $D$ mass.

Apart from the charmonium vetoes, we analyze the full
range of $q^2$. In the $B\to K e^+e^-$, $B\to K\mu^+\mu^-$, and
$B\to K^* e^+ e^-$ modes, we have good efficiency over most of the $q^2$ range,
while in the $B\to K^*\mu^+\mu^-$ mode the efficiency is highest at intermediate
and high $q^2$ values. This efficiency variation results from the interplay
between the lepton identification efficiency and the reconstruction 
efficiency of the recoil hadron as a function of momentum.
The average lepton momentum increases as $q^2$ increases;
since our minimum momentum requirement is lower for electrons than muons, the region 
of good lepton efficiency generally extends to lower values of $q^2$ for electrons. However,
the momentum of the recoil hadron decreases as $q^2$ increases, which produces
the opposite effect in the $K^{(*)}$ efficiency.

Continuum background from non-resonant $e^+e^-\to q\overline q$ production
is suppressed using a Fisher 
discriminant~\cite{bib:Fisher}, which is a linear combination of the input
variables with optimized coefficients. 
The variables included are $R_2$; $\cos \theta_B$, 
the cosine of the angle between 
the $B$ candidate and the beam axis in the CM~frame; $\cos \theta_{\rm T}$, 
the cosine of the angle between the thrust axis of the candidate $B$ meson
daughter particles and that of the remaining  
particles in the CM~frame; and 
$m_{K\ell}$, the invariant mass of the 
$K$-lepton system, 
where the lepton is selected according to
its charge relative to the strangeness 
of the $K^{(*)}$.
The variable $m_{K\ell}$ helps discriminate against background from 
semileptonic $D$ decays, for which $m_{K^{(*)}\ell}<m_D$.  
The Fisher discriminant is optimized separately for each decay channel.

Combinatorial background from $\BB$ events is suppressed 
using a signal-to-$\BB$ likelihood ratio that  
combines candidate $B$ and dilepton vertex probabilities; 
the significance of the 
dilepton separation along the beam direction; 
$\cos\theta_B$; and the missing energy, $E_{\rm miss}$, of the 
event in the CM~frame. The variable $E_{\rm miss}$ provides the
strongest discrimination against $\BB$ background, since
events with semileptonic decays usually have significant unobserved energy
due to neutrinos. We construct a likelihood variable because 
the shape of the $B$ vertex probability distributions
are not well suited to a Fisher discriminant. A separate $B$ likelihood
variable is constructed for each channel.

For each final state, we select 
at most one combination of particles per event as a $B$ signal
candidate.  If multiple candidates occur in the fit region, we select the 
candidate with the greatest
number of drift chamber and SVT hits on the charged tracks. 
In the $B\to K\ell^+\ell^-$ modes less than 1\% of the events have
multiple candidates and in the $B\to K^*\ell^+\ell^-$ modes about
10\% of the events have more than one candidate.
This criterion is designed to be negligibly biased with respect to 
the kinematic quantities used in our fit. 

\section{Control sample studies}
\label{sec:ControlSamples}

\begin{figure}[!p]
 \begin{center}
   \includegraphics[width=\linewidth]{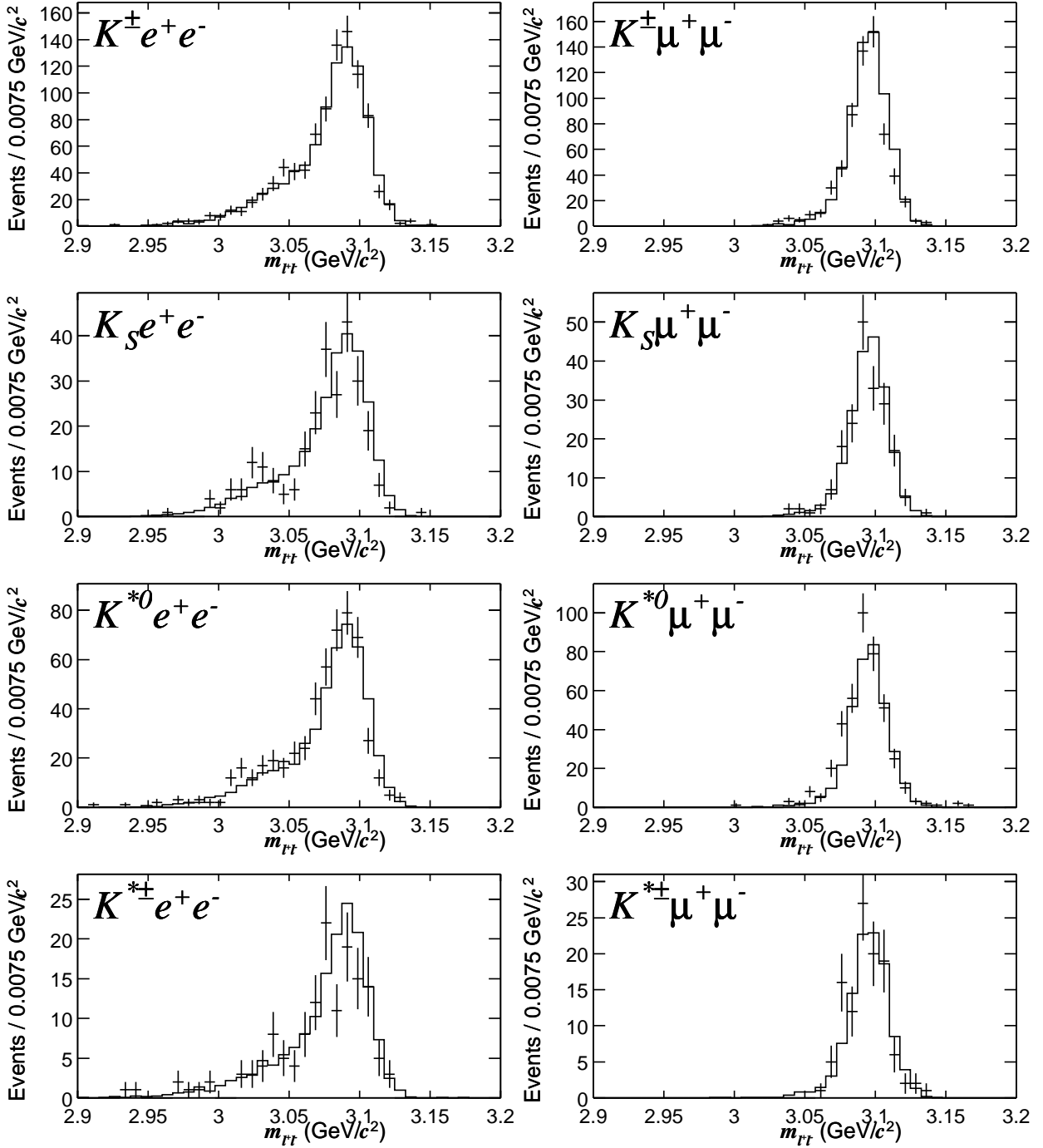}
  \end{center}
  \vspace{-0.5cm}
\caption[Comparison of $m_{\ell^+\ell^-}$ shapes in the charmonium control sample]
{\label{fig:mDilepControl}
 Comparison of the $m_{\ell^+\ell^-}$ distributions between data and Monte Carlo simulation for the charmonium control samples.
The points with error bars show the data, and the solid histograms show the 
prediction of the charmonium Monte Carlo simulation. All of the analysis selection criteria 
have been applied except for the charmonium veto, which is reversed. 
The events are also required to be in the nominal signal region in the $m_{\rm ES}$ vs.~$\Delta E$ plane.
The large tails
in the $e^+e^-$ modes are due to photon radiation. 
}
\end{figure}

\begin{figure}[!p]
 \begin{center}
   \includegraphics[width=\linewidth]{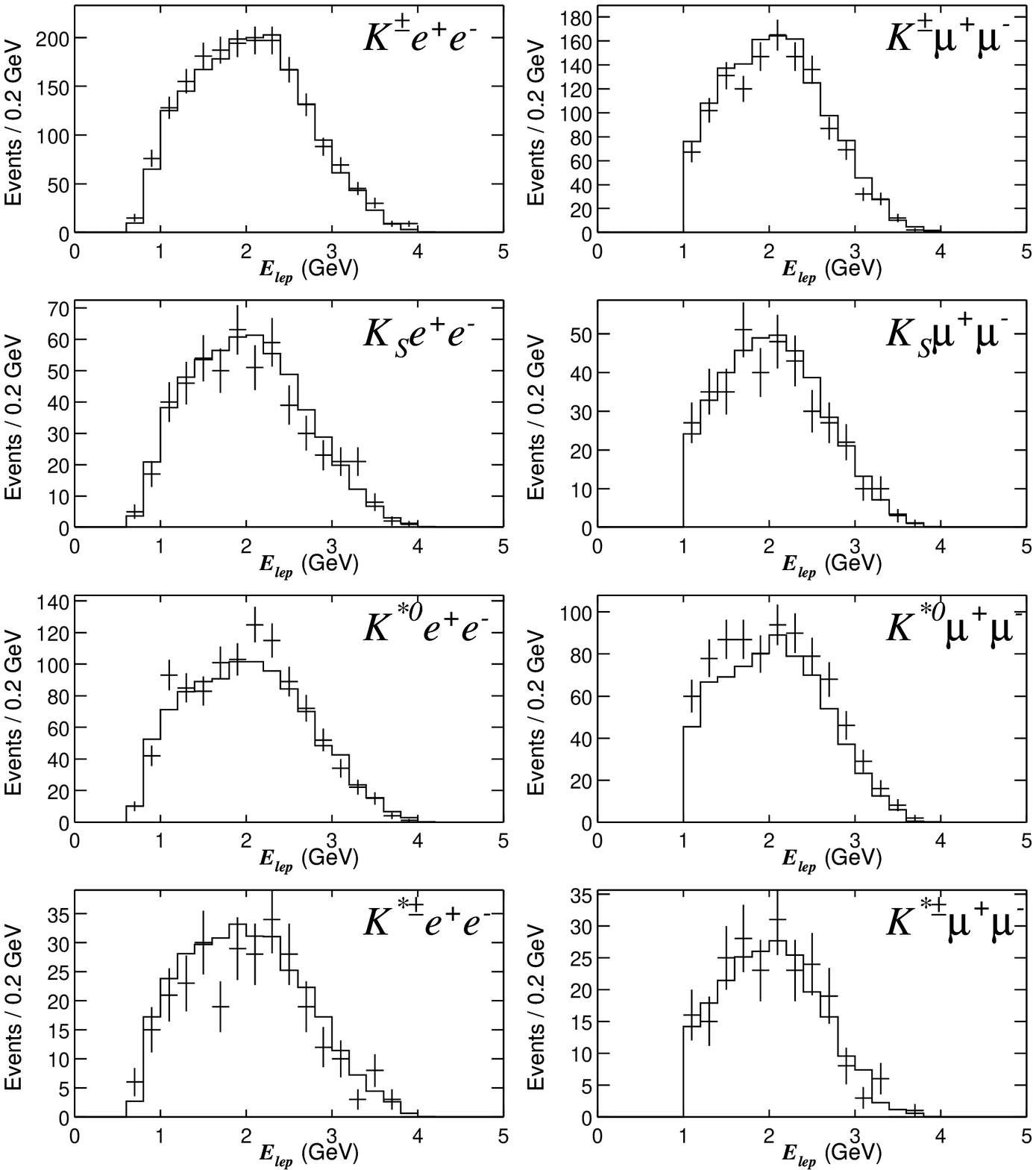}
  \end{center}
  \vspace{-0.5cm}
\caption[Comparison of lepton-energy spectra in the charmonium control sample]
{\label{fig:lepEnControl}
 Comparison of the lepton-energy distributions between data and Monte Carlo simulation for the charmonium control samples.
The points with error bars show the data, and the solid histograms show the 
prediction of the charmonium Monte Carlo simulation. All of the analysis selection criteria 
have been applied except for the charmonium veto, which is reversed. 
The  events are also required to be in the nominal signal region in the $m_{\rm ES}$ vs.~$\Delta E$ plane.
}
\end{figure}

\begin{figure}[!p]
 \begin{center}
   \includegraphics[width=\linewidth]{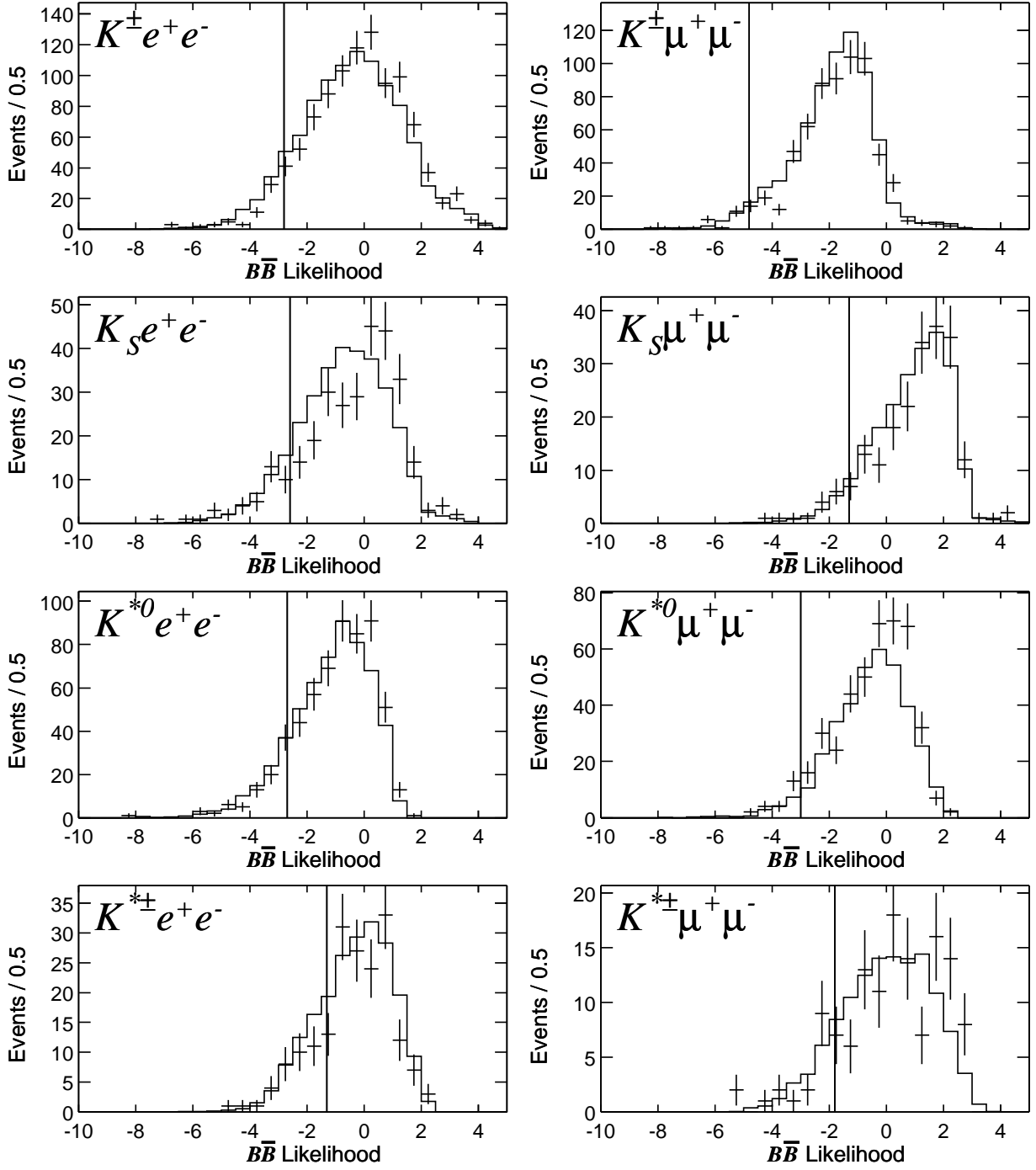}
  \end{center}
  \vspace{-0.5cm}
\caption[Comparison of the $B$ likelihood variable in the charmonium control sample]
{\label{fig:BLikelihoodControl}
 Comparison of the $B$ likelihood distributions between data and Monte Carlo simulation for the charmonium control samples.
The points with error bars show the data, and the solid histograms show the 
prediction of the charmonium Monte Carlo simulation. All of the analysis selection criteria 
have been applied except for that on the $B$ likelihood and the charmonium veto, which is reversed. 
The events are also required to be in the nominal signal region in the $m_{\rm ES}$ vs.~$\Delta E$ plane.
The vertical
lines show the minimum $B$ likelihood values allowed for the corresponding signal
modes. 
}
\end{figure}

\begin{figure}[!p]
 \begin{center}
   \includegraphics[width=\linewidth]{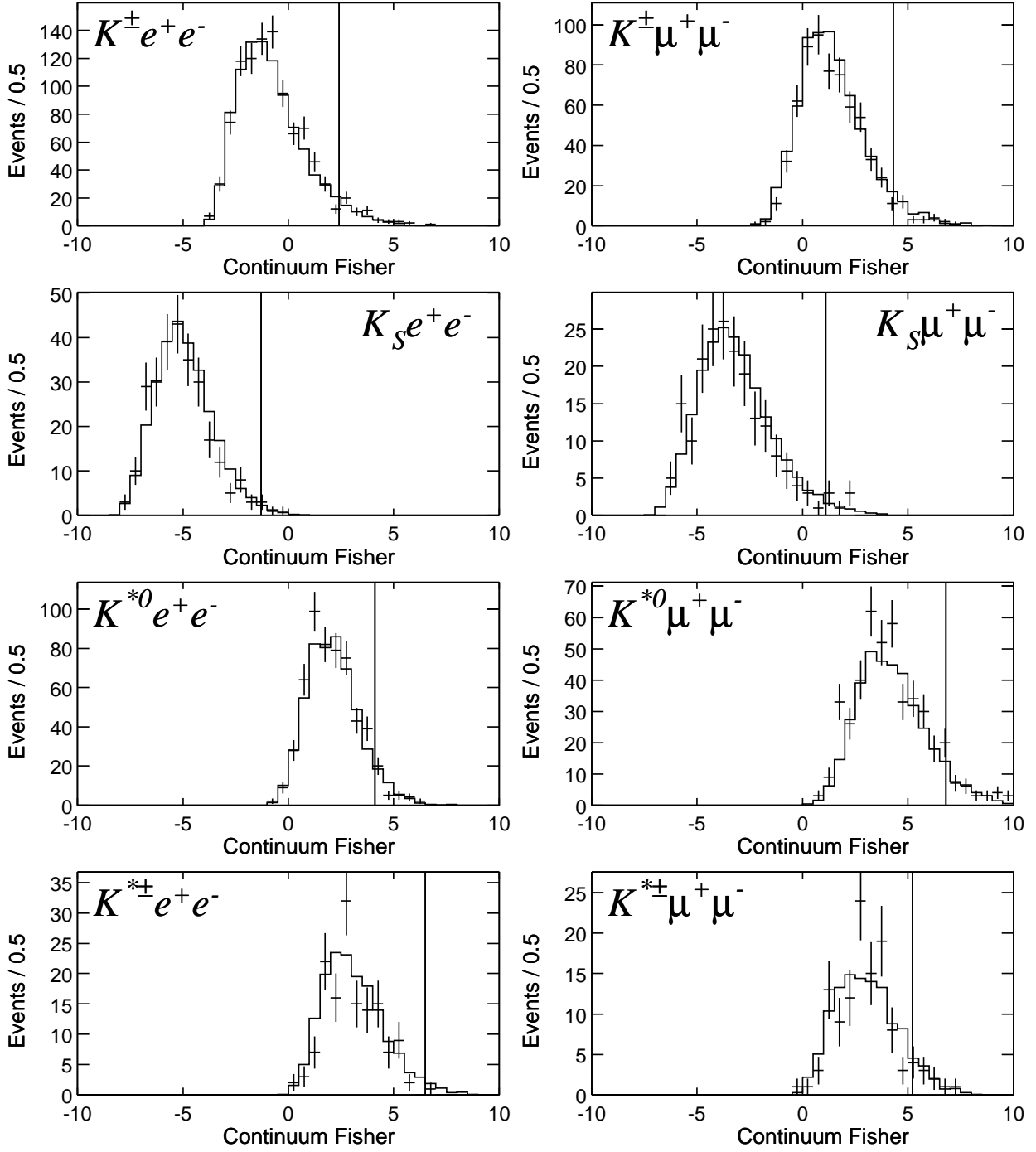}
  \end{center}
  \vspace{-0.5cm}
\caption[Comparison of the Fisher continuum-suppression variable in the charmonium control sample]
{\label{fig:FisherControl}
 Comparison of the Fisher continuum-suppression variable distributions
between data and Monte Carlo simulation for the charmonium control samples.
The points with error bars show the data, and the solid histograms show the 
prediction of the charmonium Monte Carlo simulation. All of the analysis selection criteria 
have been applied except for that on the Fisher and the charmonium veto, which is reversed. 
The events are also required to be in the nominal signal region in the $m_{\rm ES}$ vs.~$\Delta E$ plane.
The vertical
lines show the maximum Fisher values allowed for the corresponding signal
modes.

}
\end{figure}

To check the
ability of the Monte Carlo to correctly simulate the
detector response, we study the charmonium decays 
$B\to J/\psi K^{(*)}$ and $B \to \psi(2S) K^{(*)}$, whose
decay distributions and branching fractions are reasonably well known.
Figures~\ref{fig:mDilepControl}, \ref{fig:lepEnControl}, \ref{fig:BLikelihoodControl},
and \ref{fig:FisherControl} compare the distributions of
$m_{\ell^+\ell^-}$, lepton energy in the $\Upsilon(4S)$ CM~frame, 
$B$ likelihood variable, and Fisher continuum suppression variable for
these signal-like control samples. In each case the distributions
are absolutely normalized, so that we are comparing not only the
agreement between shapes, but also the agreement between event yields.  
Table~\ref{tab:CharmoniumYields} summarizes the event yields in the
nominal signal boxes for data and the Monte Carlo prediction in the control 
samples. We observe good agreement in the yield between data and Monte Carlo
simulation in
all channels. The agreement in the tails of the dilepton mass
distributions indicates that the placement of our charmonium
vetoes is based on a reliable simulation.
The lepton-energy distributions are well simulated,
indicating that the Monte Carlo efficiencies should be well
simulated at other values of $q^2$. The selections 
on the $B$
likelihood and Fisher continuum-suppression variables
are rather loose, and the agreement between Monte Carlo simulation
and data in the control samples indicates that the
Monte Carlo efficiencies should be reliable. We discuss
these issues quantitatively in the discussion of systematic
errors (Sec.~\ref{sec:Systematics}).

 
\begin{table}[ht]
 \caption[Summary of \jpsi yields in data and Monte Carlo.]{ 
  Comparison of the \jpsi control sample event yields in data and Monte
  Carlo simulations for each of the modes.  All selection 
  criteria have been applied,
  except that the \jpsi veto is reversed, and the nominal signal region is
  selected in \mes and \DeltaE.  The Monte Carlo samples have been
  scaled to the equivalent number of \FourS decays in the corresponding
  data sample.  Errors on the ratios of data yields to yields from 
  the Monte Carlo
  simulation are
  due to our sample statistics only, and do not incorporate branching
  fraction uncertainties.
 }
 \begin{center}
 \begin{tabular}{lrrr}\hline\hline
  {Mode ($J/\psi\to\ell^+\ell^-)$} & {MC Yield} 
	& {Data Yield} & {Data/MC}(\%) \\ 
  \hline
  \modekee     & 911.8 & 939 &  ${103.0}{\pm 3.8}$ \\    	           	      	    	                 	
  \modekmm     & 669.1 & 623 &  ${ 93.1}{\pm 4.1}$ \\  
  \modeksee    & 271.6 & 265 &  ${ 97.6}{\pm 6.0}$ \\       	      	    	                 	
  \modeksmm    & 195.0 & 191 &  ${ 98.0}{\pm 7.1}$ \\ \hline    	           	      	    	                 	
  \modekstkee   & 494.5 & 523 & $ {105.8}{\pm 4.9}$ \\      	           	      	    	                 	
  \modekstkmm   & 353.7 & 412 &  ${116.5}{\pm 6.2}$ \\            	      	    	                 	
  \modekstksee  & 158.6 & 144 &  ${ 90.8}{\pm 7.7}$ \\              	      	    	                 	
  \modekstksmm  & 106.4 & 111 &  ${104.4}{\pm 10.2}$ \\ \hline 
  \multicolumn{4}{l}{Combined by hadron final state} \\ 
  \modekll     & 1580.9 & 1562 &  ${ 98.8}{\pm 2.8}$ \\  
  \modeksll    &  466.5 &  456 &  ${ 97.7}{\pm 4.6}$ \\  
  \modekstkll  &  848.2 &  935 &  ${110.2}{\pm 3.9}$ \\  
  \modekstksll &  264.9 &  255 &  ${ 96.3}{\pm 6.2}$ \\  
   \hline
  \multicolumn{4}{l}{Combined by lepton final state} \\ 
  \epem modes  & 1836.4 & 1871 &  ${101.9}{\pm 2.6}$ \\  
  \mumu modes  & 1324.1 & 1337 &  ${101.0}{\pm 3.0}$ \\  
  \hline  all modes    & 3160.5 & 3208 &  ${101.5}{\pm 1.9}$ \\ \hline\hline 
 \end{tabular}
 \end{center}

 \label{tab:CharmoniumYields}
\end{table}

\begin{figure}[!p]
 \begin{center}
   \includegraphics[width=\linewidth]{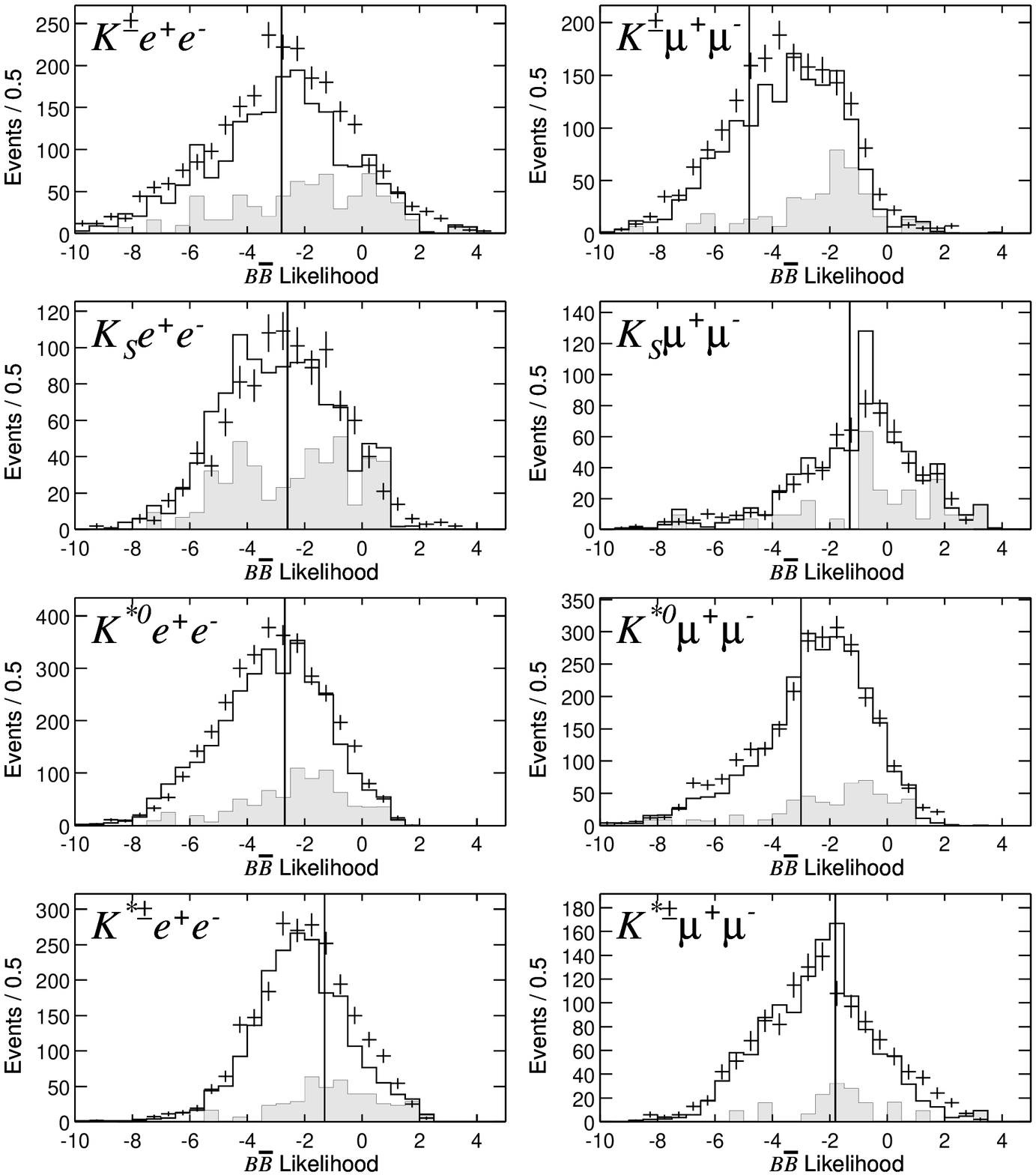}
  \end{center}
  \vspace{-0.5cm}
\caption[Comparison of the $B$ Likelihood variable in the large sideband sample]
{\label{fig:sidebandBLikelihood}
Large sideband comparison of the $B$ likelihood variable distributions 
between data and $B\overline B$ Monte Carlo simulation plus scaled off-resonance data.
The points with error bars show the on-resonance data, the shaded histograms
show the scaled off-resonance data, and the open histograms show the sum
of the scaled off-resonance data and the $B\overline B$ Monte Carlo sample.
All of the analysis selection criteria 
have been applied except for that on the $B\overline B$ likelihood. The
position of this cut is indicated by vertical lines; selected
events are required to have $B$ likelihood above this value. 
}
\end{figure}

The large sidebands provide a useful tool for monitoring our
simulation of the combinatorial backgrounds. 
In contrast to the charmonium control samples, the events here are produced
by a wide variety of processes, including both $B\overline B$ and continuum.
We compare the distributions observed in data with the sum of $B\overline B$
Monte Carlo simulations and continuum samples obtained from off-resonance
data. Although the off-resonance sample is statistically limited, it gives
a correct picture of the complicated mix of $q\overline q$, $\tau^+\tau^-$,
and two-photon events. 
Figure~\ref{fig:sidebandBLikelihood}
shows the distribution of the $B$ likelihood variable in the large sideband.
Both the shapes and the normalizations are reasonably well predicted.
There are large fluctuations in the off-resonance data sample, which is
scaled by a factor of about 8 to correspond to the integrated
luminosity of the on-resonance data sample. 

\begin{figure}[!p]
 \begin{center}
   \includegraphics[width=\linewidth]{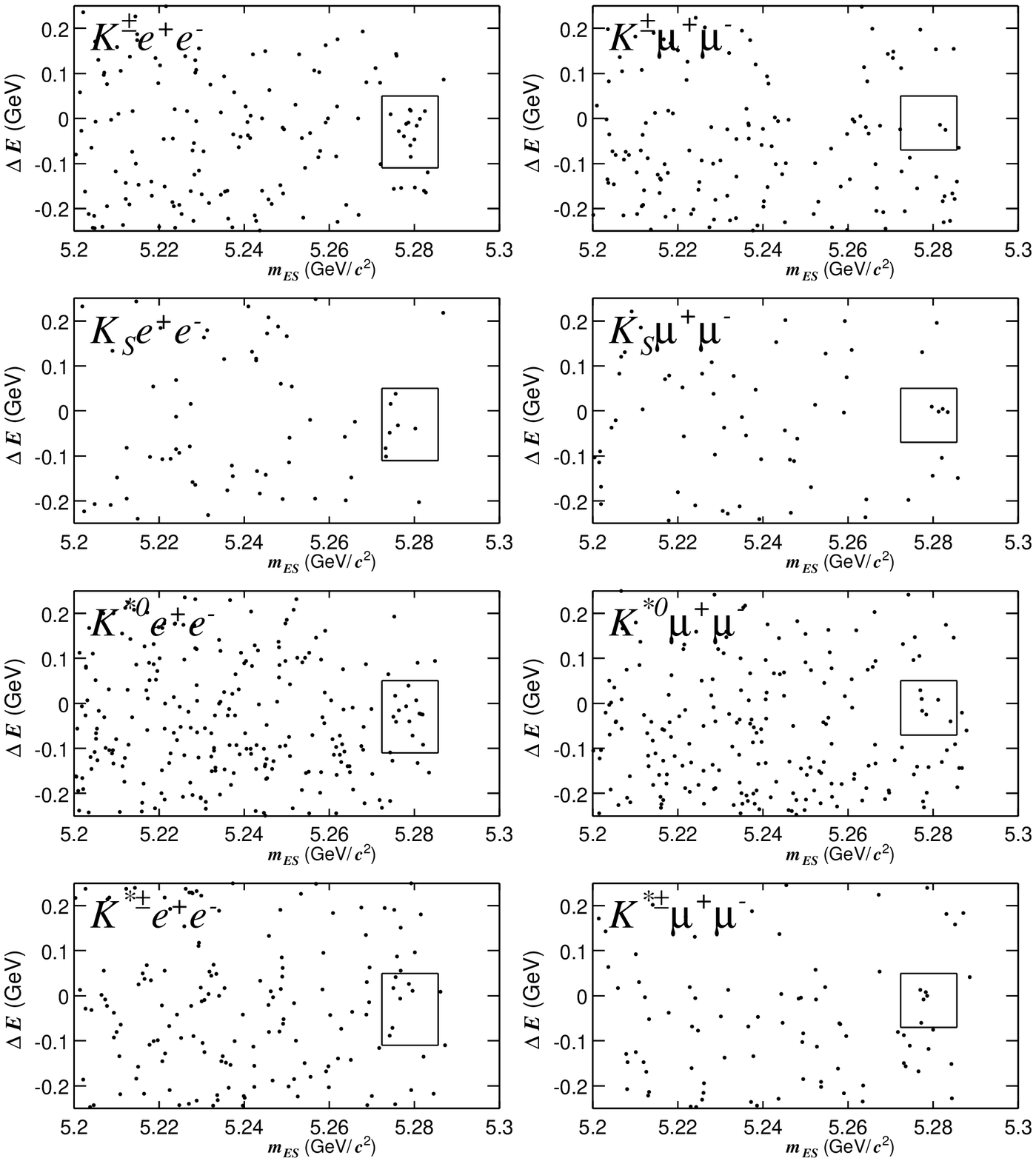}
  \end{center}
  \vspace{-0.5cm}
\caption[Data: scatterplots in $m_{\rm ES}$ vs.~$\Delta E$.]
{\label{fig:DataScatterplots}
Distribution of the data in the $\Delta E$ vs.~$m_{\rm ES}$ plane 
(fit region) for each channel. The rectangles show the nominal
signal boxes, but the signals are extracted from a two-dimensional 
fit.}
\end{figure}

\begin{figure}[!p]
 \begin{center}
   \includegraphics[width=\linewidth]{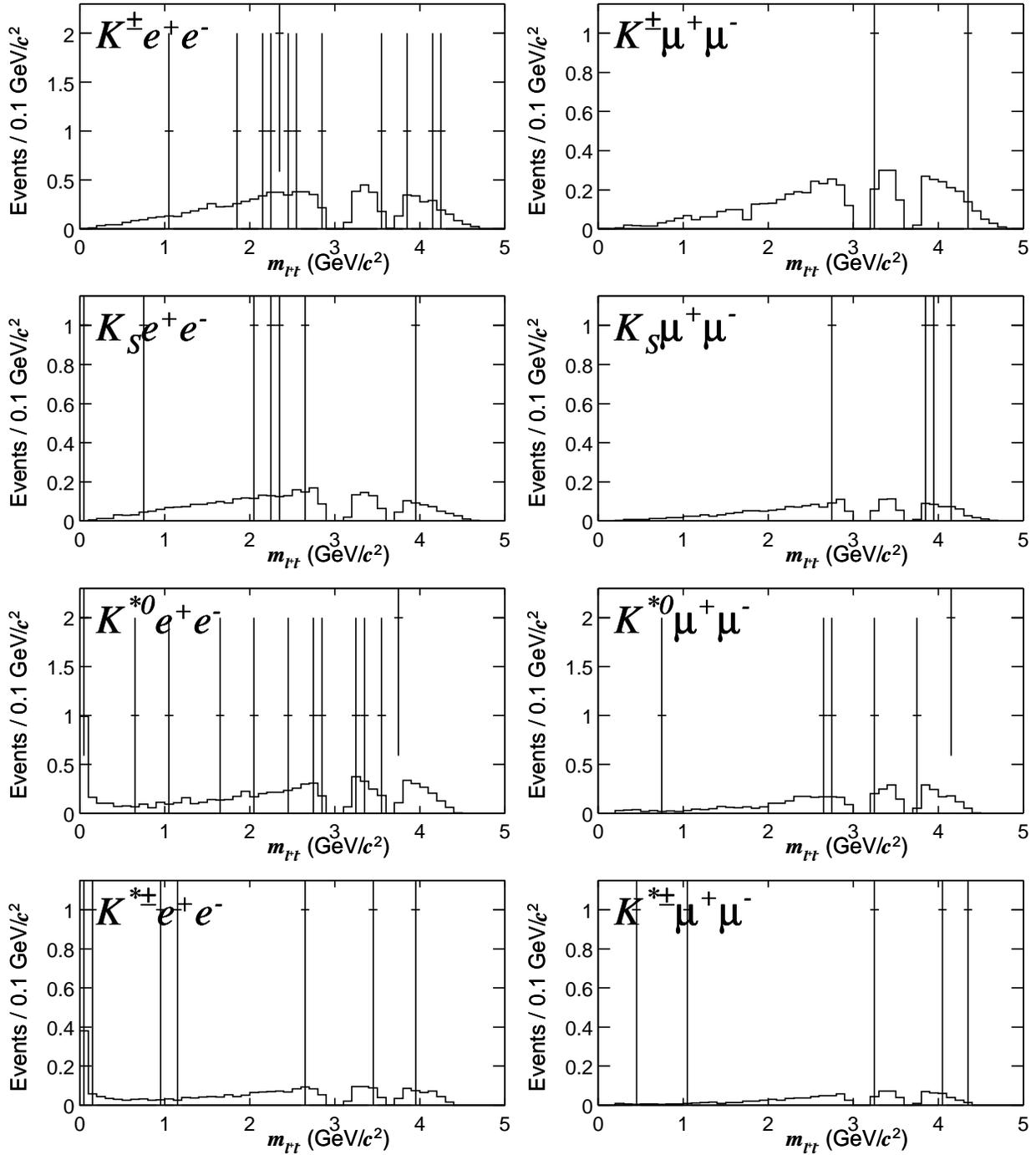}
  \end{center}
  \vspace{-0.5cm}
\caption[]
{\label{fig:dileptonmass}
The dilepton mass distribution for data (points with error bars)
and signal Monte Carlo (histogram) for events in the nominal
signal region. The data include some amount of background,
so the distributions are not expected to agree perfectly. However,
the data do not cluster around the charmonium veto regions; such
clustering would be a possible indication of charmonium 
background events escaping the
veto.
}
\end{figure}

For each data taking period, we check the control
samples in the manner described above.
We then ``unblind'' the data in the fit region. 
The distributions of the data for each mode 
are shown in Fig.~\ref{fig:DataScatterplots}. In several
of the modes, a small number of events is present
in the nominal signal region. For the events in the nominal
signal region we show the dilepton mass distribution in 
Fig.~\ref{fig:dileptonmass}. The prediction based on our
simulation of the signal is also shown.

\section{Fits and yields}
\label{sec:FitsAndYields}

We extract the signal and background yields in each channel
using a two-dimensional extended unbinned maximum 
likelihood fit in the region 
defined by \mbox{$m_{\rm ES}$ $>$ 5.2 GeV/$c^2$} and
\mbox{$|\Delta E|$ $<$ 0.25 GeV} (``fit region''). 
The two-dimensional signal shapes,
including the effects of radiation on the $\Delta E$
distribution and the correlation between $m_{\rm ES}$
and $\Delta E$,  
are obtained by parameterizing a
{\tt GEANT}4 Monte Carlo simulation of the signal
using the product of two Crystal Ball shapes~\cite{bib:CBShape},
one for $m_{\rm ES}$ and one for $\Delta E$.
The Crystal Ball shape
smoothly matches a Gaussian core with a tail that is used to describe
radiative effects. For $m_{\rm ES}$ we use a single Gaussian core, while
in $\Delta E$ we describe the core using the 
sum of two Gaussians with separate mean and width. The parameters in 
the Crystal Ball shape for $m_{\rm ES}$ are allowed to have a 
quadratic dependence on $\Delta E$, which adequately takes
into account the correlation between $m_{\rm ES}$ and $\Delta E$.
Small shifts in the mean values of $m_{\rm ES}$ and $\Delta E$ 
are applied to the signal shapes based on studies of these
quantities in the charmonium control samples.

The combinatorial background is described by the function

\begin{equation}
f(\DeltaE,\mes)=Ne^{s\Delta E}\mes\sqrt{1-{\mes^2\over E_{\rm b}^2}}
                 e^{-\xi\left(1-{m_{\rm ES}^2\over E_{\rm b}^2}\right)},
\nonumber
\end{equation}
where $N$ is a normalization factor, and $s$ and $\xi$ are free
parameters determined in the fit to the data. The part of this
function that describes the $m_{\rm ES}$ distribution is
a standard form known as the ARGUS function~\cite{bib:ArgusFunction}.

For systematic studies we also use a more general form for the
background shape, where the ARGUS slope parameter, $\xi$, is allowed to be a
function of $\Delta E$:
\begin{equation}
\xi=\xi(\Delta E)=\xi_0+\xi_1\Delta E+{1\over2}\xi_2(\Delta E)^2.\nonumber
\label{eq:xi}
\end{equation}

In each channel we also include a term to describe
the estimated amount of background that peaks under the signal.
Peaking backgrounds
can arise from $B\to K^*\gamma$ followed by a photon
conversion in the detector material or from decays such
as $B\to K\pi\pi$ in which the pions are misidentified as leptons.
Our estimates for these backgrounds are obtained from specific
studies of individual processes and make use of hadron misidentification
rates measured in data. The total estimated peaking background 
for each mode is listed in Table~\ref{tab:peakingbkgd}.
\begin{table}[ht] 
 \caption[Summary of peaking backgrounds used in the fits.]
 {Summary of peaking backgrounds used in the fits.} 
\begin{center} 
\begin{tabular}{p{2in}l} \hline\hline
{Mode} & 
\mbox{\hspace{-4em}Peaking background (events)} \\ 
\hline \vspace*{-10pt} 
 \modekee     & $0.0^{+1.0}_{-0.0}$ \\ 
 \modekmm     & $0.7\pm0.7$ \\ 
 \modeksee    & $0.0^{+0.1}_{-0.0}$ \\ 
 \modeksmm    & $0.5\pm0.5$ \\ 
 \modekstkee  & $0.2^{+0.5}_{-0.2}$ \\ 
 \modekstkmm  & $1.2\pm1.2$ \\
 \modekstksee & $0.1^{+0.4}_{-0.1}$ \\ 
 \modekstksmm & $1.0\pm1.0$ \\ \hline\hline
\end{tabular} 
\end{center}
\label{tab:peakingbkgd}
\end{table}

\begin{figure}[!p]
 \begin{center}
   \includegraphics[width=\linewidth]{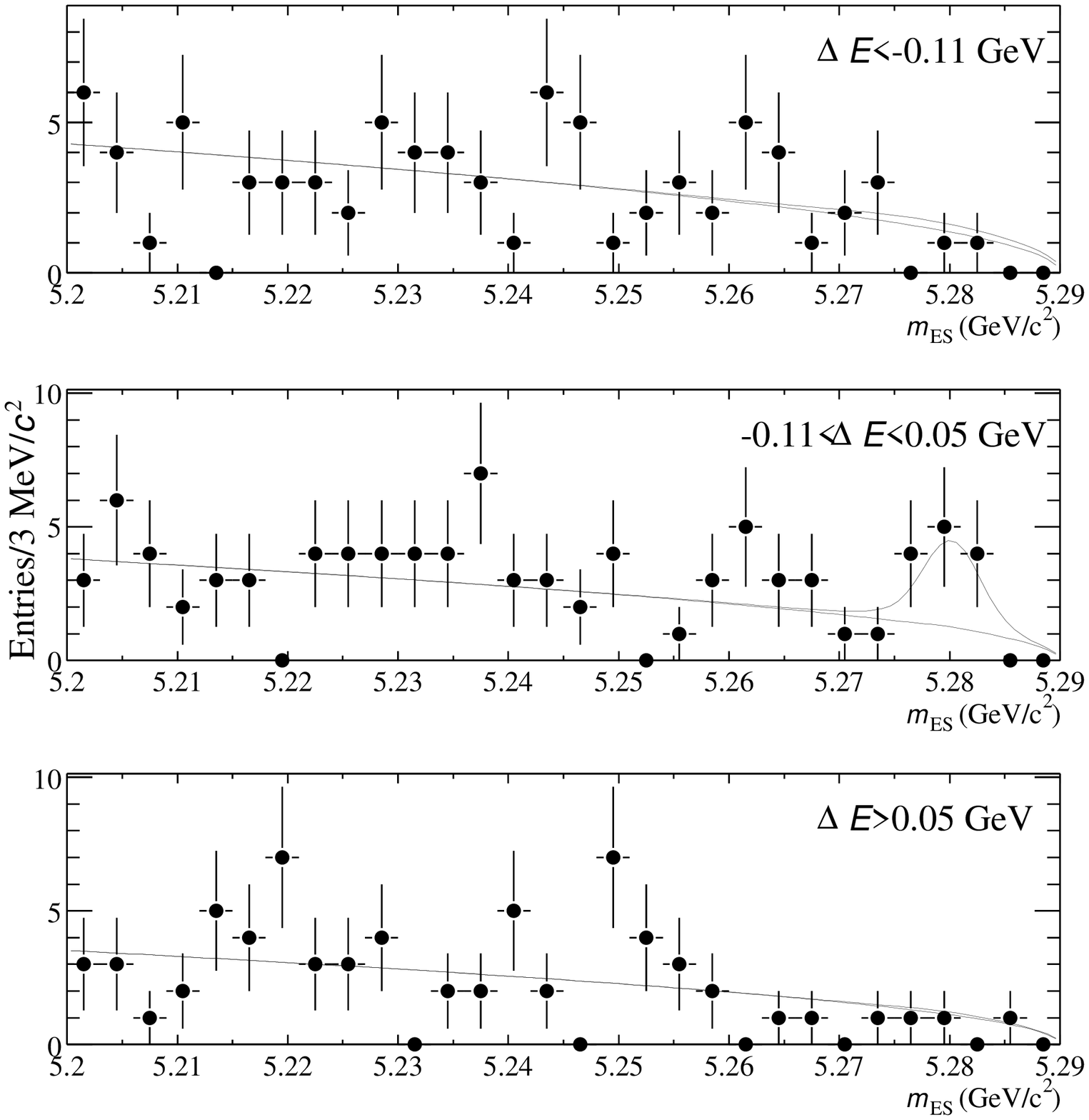}
  \end{center}
  \vspace{-0.5cm}
\caption[Kstkee mES fits]
{\label{fig:kstkee_mES}
Fit to the data for the mode $B^0\to K^{*0}(\to K^+\pi^-)e^+e^-$.
The three histograms show the $m_{\rm ES}$ distributions in three
slices of $\Delta E$:  $\Delta E<-0.11$ GeV, 
$-0.11\le\Delta E<0.05$ GeV, and $\Delta E\ge0.05$ GeV. 
}
\end{figure}

\begin{figure}[!p]
 \begin{center}
   \includegraphics[width=\linewidth]{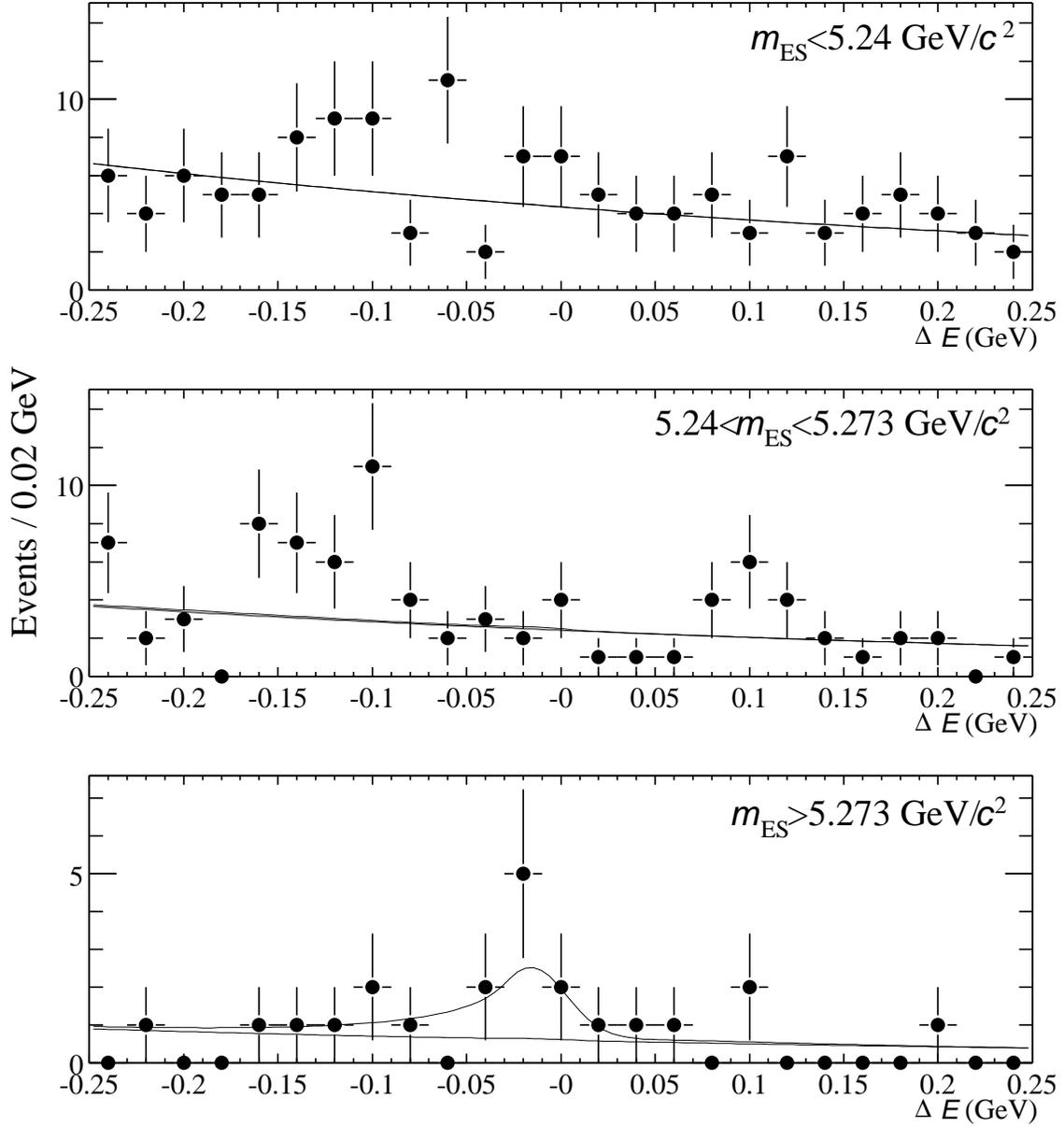}
  \end{center}
  \vspace{-0.5cm}
\caption[Kstkee Delta E fits]
{\label{fig:kstkee_DeltaE}
Fit to the data for the mode $B^0\to K^{*0}(\to K^+\pi^-)e^+e^-$.
The three histograms show the $\Delta E$ distributions in three
slices of $m_{\rm ES}$: $m_{\rm ES}<5.24$ GeV$/c^2$, $5.24\le m_{\rm ES}<5.273$ GeV$/c^2$,
$m_{\rm ES}\ge 5.273$ GeV$/c^2$. The peak in
$\Delta E$ near zero in the lower histogram is correlated with the
peak in $m_{\rm ES}$ around the $B$ mass. 
}
\end{figure}

As an example, we present the fits for the $B\to K^{*0}e^+e^-$ mode.
Figure~\ref{fig:kstkee_mES} shows the $m_{\rm ES}$ distributions in three
slices of $\Delta E$ that span the fit region: $\Delta E<0.11$ GeV, 
$-0.11\le\Delta E<0.05$ GeV, and $\Delta E\ge0.05$ GeV. We observe
a peak in $m_{\rm ES}$ around the $B$ mass for the range
$-0.11\le\Delta E<0.05$ GeV, but not in the upper or lower $\Delta E$
sidebands. Figure~\ref{fig:kstkee_DeltaE} shows the distributions of
the data in $\Delta E$ for three slices of $m_{\rm ES}$: 
$m_{\rm ES}<5.24$ GeV$/c^2$, $5.24\le m_{\rm ES}<5.273$ GeV$/c^2$,
$m_{\rm ES}\ge 5.273$ GeV$/c^2$. 

\begin{figure}[!p]
 \begin{center}
   \includegraphics[width=\linewidth]{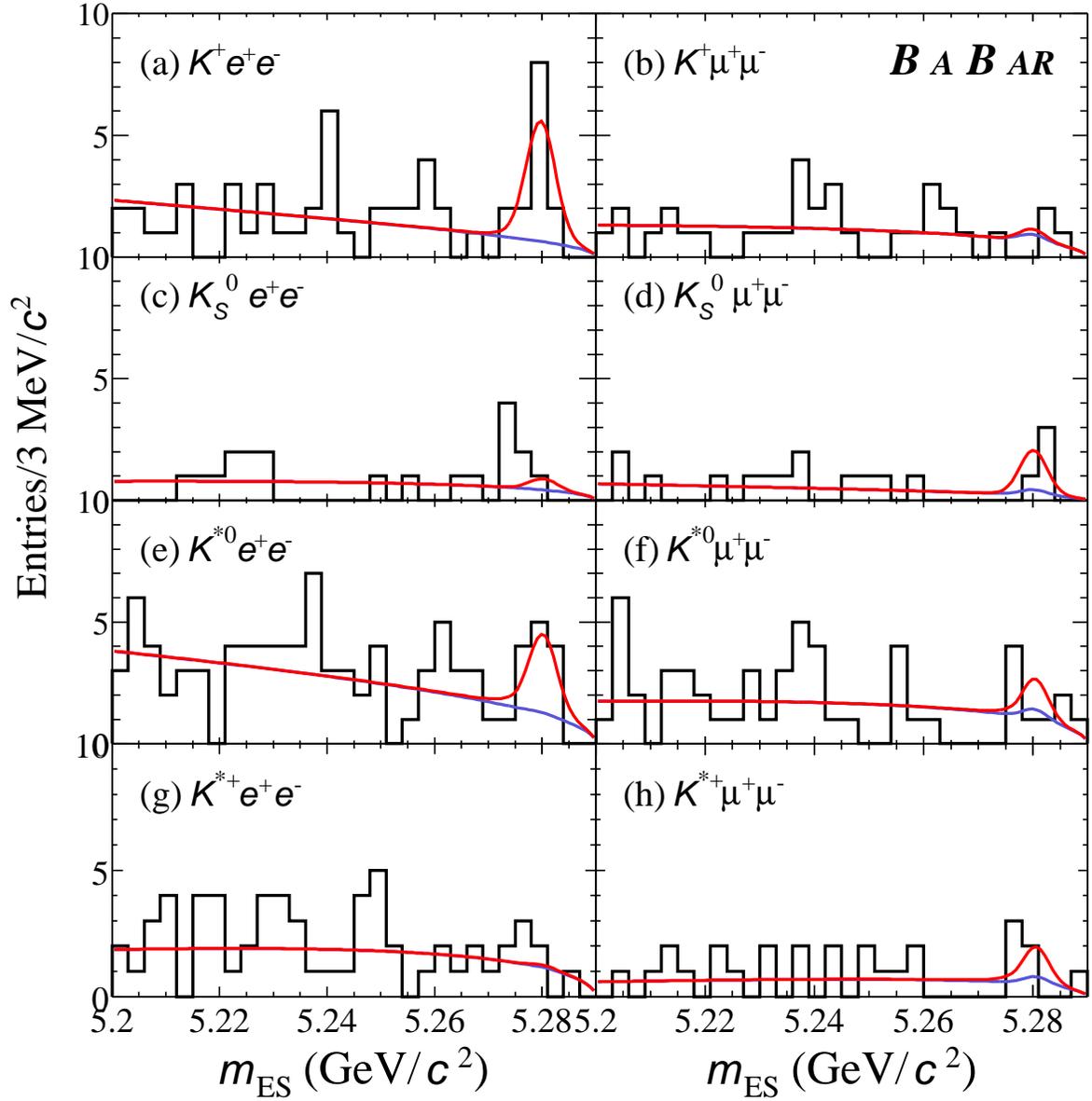}
  \end{center}
  \vspace{-0.5cm}
\caption[Data: projections of fits on $m_{\rm}$]
{\label{fig:DataFitsMES}
Projections of the individual fits onto the $m_{\rm ES}$ distributions for
the slice $-0.11\le\Delta E<0.05$ GeV (electron modes) or
$-0.07\le\Delta E<0.05$ GeV (muon modes).
}
\end{figure}

\begin{figure}[!p]
 \begin{center}
   \includegraphics[width=\linewidth]{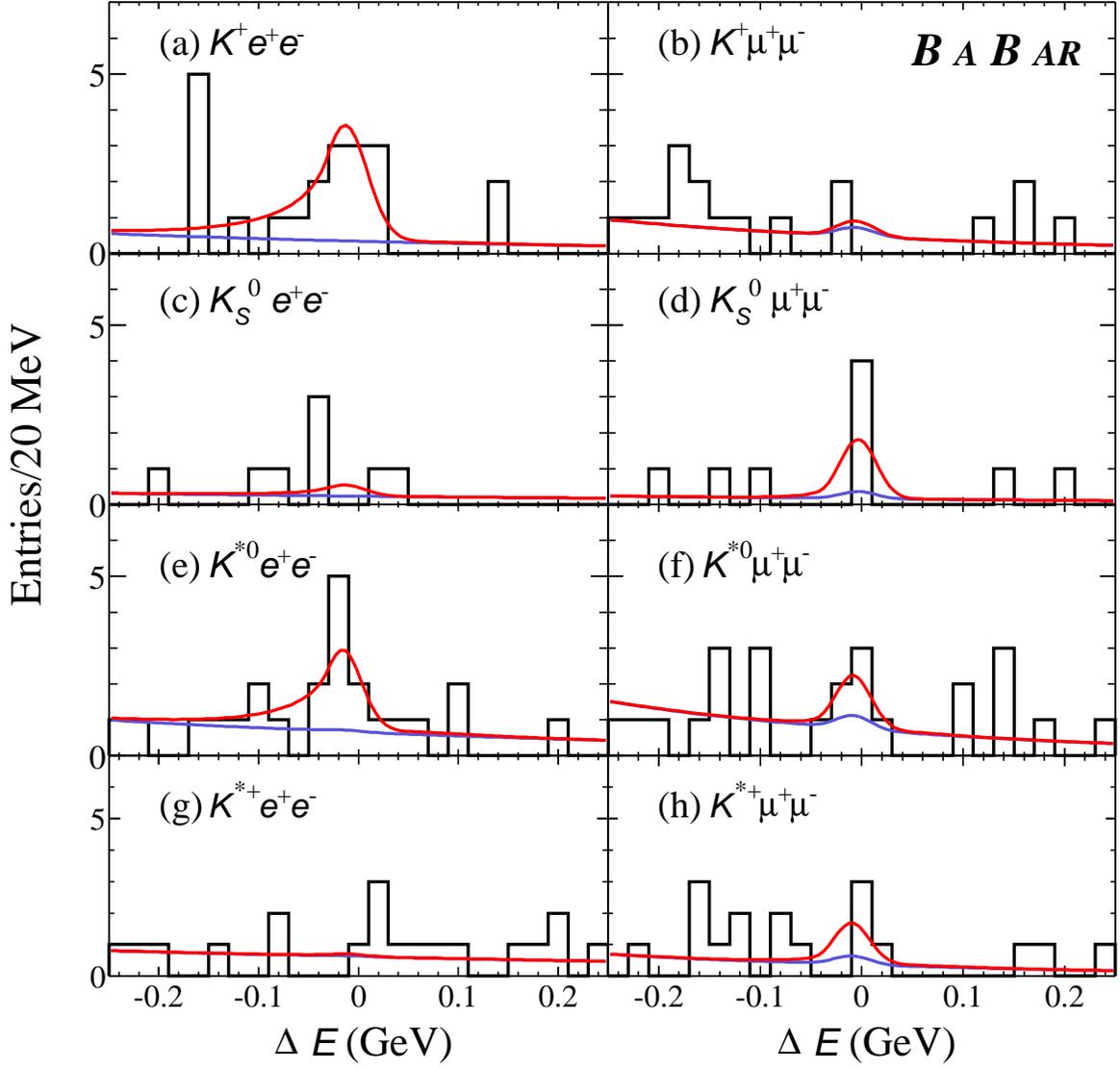}
  \end{center}
  \vspace{-0.5cm}
\caption[Data: projections of fits on $\Delta E$.]
{\label{fig:DataFitsDeltaE}
Projections of the individual fits onto the $\Delta E$ distributions for
$m_{\rm ES}$ in the nominal signal regions.
}
\end{figure}

Figure~\ref{fig:DataFitsMES} shows the projections of
the fits for each channel onto $m_{\rm ES}$ for 
$\Delta E$ in the nominal signal regions. 
Figure~\ref{fig:DataFitsDeltaE} shows the
projections of the fits for each channel onto
$\Delta E$ for $m_{\rm ES}$ in the nominal
signal regions. We observe small excesses
of events over the backgrounds in several
channels. Table~\ref{tab:results} summarizes the results for each
channel. 

\begin{table}
  \caption[Fit results]{ Results from individual fits to \kll modes.  The
  columns are, from left to right: (1) the fitted signal yield;
  (2) the contribution of the background to the error on the signal yield,
  expressed as an effective background yield; (3) the signal efficiency,
  $\epsilon$ (not including the branching fractions for \Kstar, \Kz,
  and \KS decays); (4) the multiplicative systematic error on the selection efficiency,
  $(\Delta\BR/\BR)_\epsilon$; (5) the additive systematic error from the fit,
  $(\Delta\BR)_{\rm fit}$; and (6) the branching fraction central value
  (\BR).  } \label{tab:results}
 \begin{center}
 \begin{tabular}{lrccccl}
\hline\hline
Mode & Signal & Effective & $\epsilon$  & ($\Delta {\cal B}/{\cal B})_{\epsilon}$   & $(\Delta{\cal B})_{\rm fit}$ &
$\qquad{\cal B}$ \\
& Yield  & Bkgnd.    &  (\%)      & (\%)   & $(/10^{-6})$ & \ \ $(/10^{-6})$  \\ 
 \hline 
$B^+\to K^+e^+e^-$                         &  $14.4^{+5.0}_{-4.2}$    &   2.2  & 17.5       &   $\pm 6.8$ 
             &  $^{+0.14}_{-0.21}$         &  $0.98^{+0.34+0.16}_{-0.28-0.22}$  \\    
$B^+\to K^+\mu^+\mu^-$                     &  $0.5^{+2.3}_{-1.3}$     &   2.2  & 9.2        &   $\pm 6.6$ 
             &  $^{+0.09}_{-0.08}$         &  $0.06^{+0.30+0.09}_{-0.17-0.08}$  \\    
$B^0\to K^0 e^+e^-$                        &  $1.3^{+2.6}_{-1.7}$     &   1.5  & 18.6       &   $\pm 7.9$  
            &  $\pm 0.14$      &  $0.24^{+0.49+0.14}_{-0.32-0.15}$  \\    
$B^0\to K^0 \mu^+\mu^-$                    &  $3.6^{+2.9}_{-2.1}$     &   0.9  & 9.4        &   $\pm 7.7$ 
             &  $\pm 0.24$      &  $1.33^{+1.07}_{-0.78}\pm0.26$  \\    
$B^0\to K^{*0}e^+e^-$                      &  $10.6^{+5.2}_{-4.3}$    &   5.5  & 10.6       &   $\pm 7.6$
              &  $^{+0.46}_{-0.47}$      &  $1.78^{+0.87+0.48}_{-0.72-0.49}$  \\    
$B^0\to K^{*0}\mu^+\mu^-$                  &  $3.4^{+3.9}_{-2.8}$     &   4.6  &  6.1       &   $\pm 9.3$  
            &  $\pm 0.38$      &  $0.99^{+1.14}_{-0.82}\pm0.39$  \\    
$B^+\to K^{*+}e^+e^-$                      &  $0.3^{+3.7}_{-2.3}$     &   6.3  & 10.3       &   $\pm 9.5$ 
             &  $^{+0.69}_{-0.72}$   &  $0.15^{+1.87+0.69}_{-1.16-0.72}$  \\    
$B^+\to K^{*+}\mu^+\mu^-$                  &  $3.6^{+3.9}_{-2.5}$     &   3.4  &  5.2       &   $\pm 11.1$ 
            &  $\pm 1.80$      &  $3.61^{+3.91}_{-2.51}\pm1.84$  \\ \hline\hline    
\end{tabular}
\end{center}
\end{table}

\begin{figure}[!p]
 \begin{center}
   \includegraphics[width=\linewidth]{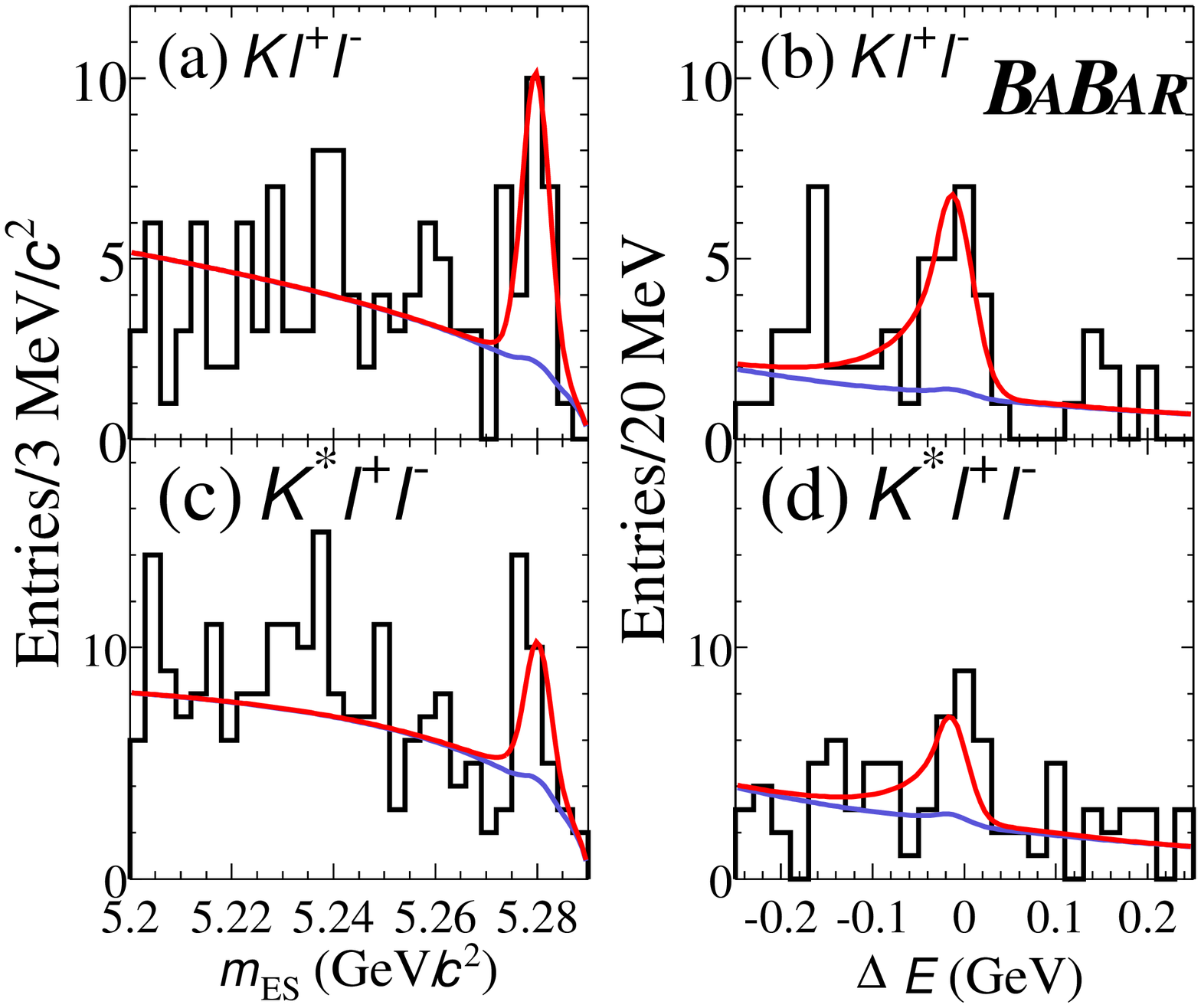}
  \end{center}
  \vspace{-0.5cm}
\caption[Data: combined fits]
{\label{fig:DataCombinedFits}
Summed distributions for the four $B\to K\ell^+\ell^-$ and the
four $B\to K^*\ell^+\ell^-$ channels, showing the projections
of the combined fit. In this combined fit, the yields in the 
$B\to K\ell^+\ell^-$ channels are constrained to each other, 
as are the yields in the $B\to K^*\ell^+\ell^-$ channels (see text).
}
\end{figure}

We also perform combined fits in which the branching
fractions in the four $B\to K\ell^+\ell^-$ channels
are constrained to each other, as are the branching fractions
for the four $B\to K^*\ell^+\ell^-$ channels 
(Fig.~\ref{fig:DataCombinedFits}). 
Due to the larger pole at low values of $m_{\ell^+\ell^-}$ 
in $B\to K^*e^+e^-$ than in $B\to K^*\mu^+\mu^-$, the expected
rates for these channels differ and are constrained to
the ratio of branching fractions 
${\cal B}(B\to K^*e^+e^-)/{\cal B}(B\to K^*\mu^+\mu^-)=1.2$ from the model
of Ali {\it et al.}~\cite{bib:TheoryA}.
The extracted yield corresponds to the electron mode. The branching
fractions from the combined fits, with statistical errors only, are
\begin{eqnarray}
\BR(B\to K\ell^+\ell^-)&=&(0.78^{+0.24}_{-0.20}(\rm stat.))\times 10^{-6},\nonumber\\
\BR(B\to K^*\ell^+\ell^-)&=&(1.68^{+0.68}_{-0.58}(\rm stat.))\times 10^{-6}.\nonumber
\end{eqnarray}

\section{Systematic uncertainties}
\label{sec:Systematics}

Systematic errors arise from uncertainties on the efficiency and on
the number of $B\overline B$ pairs in the data sample (both
of which are multiplicative errors) and on the yields
extracted from the fits (additive systematic errors).

Table~\ref{tab:systematic_mult} summarizes the multiplicative
systematic uncertainties.
The systematic uncertainty from the continuum Fisher and $\BB$
likelihood selection criteria are determined from the differences of 
efficiency for
these criteria in data and Monte Carlo simulations for charmonium control
samples (see Figures~\ref{fig:BLikelihoodControl} and
\ref{fig:FisherControl}).
The uncertainties on the efficiencies due to model-dependence of 
form factors are taken to be the full range of variation obtained 
from different theoretical models~\cite{bib:TheoryA}. 
The total multiplicative systematic uncertainty is the sum in
quadrature of the individual sources, with the exception of the
tracking efficiency uncertainties for leptons and hadrons, which are
taken to be 100\% correlated.  The total uncertainty ranges from
7--11\%; the largest individual source is the theoretical model
dependence of the signal efficiency, which ranges from 4--7\%.
The combined multiplicative systematic
errors are also listed in Table~\ref{tab:results}. 

\begin{table}[h]
\caption[Multiplicative systematic uncertainties]
 {Multiplicative systematic uncertainties.}
\footnotesize 
\begin{center} 
\begin{tabular}{lcccccccc}
\hline\hline
{Systematic} & 
$K^+e^+e^-$ & $K^+\mu^+\mu^-$ & $K_S^0e^+e^-$ & $K_S^0\mu^+\mu^-$ & 
$K^{*0}e^+e^-$ & $K^{*0}\mu^+\mu^-$ & $K^{*+}e^+e^-$ & $K^{*+}\mu^+\mu^-$ 
\\ \hline 
Trk eff. ($e,\mu$) & 
$\pm1.6$ & $\pm1.6$ & $\pm1.6$ & $\pm1.6$ & 
$\pm1.6$ & $\pm1.6$ & $\pm1.6$ & $\pm1.6$ \\ 
Electron ID & 
$\pm2.7$ & - & $\pm2.7$ & - & 
$\pm2.7$ & - & $\pm2.7$ & - \\ 
Muon ID &
 - & $\pm2.0$ & - & $\pm2.0$ &
 - & $\pm2.0$ & - & $\pm2.0$ \\ 
$K,\pi$ ID & 
$\pm2.0$ & $\pm2.0$ & - & - &
$\pm4.0$ & $\pm4.0$ & $\pm2.0$ & $\pm2.0$ \\ 
Trk eff. ($K,\pi$) & 
$\pm1.3$ & $\pm1.3$ & $\pm2.6$ & $\pm2.6$ &
$\pm2.6$ & $\pm2.6$ & $\pm3.9$ & $\pm3.9$ \\ 
$K_S^0$ eff.  &
 - & - & $\pm3.2$ & $\pm3.2$ &
 - & - & $\pm3.2$ & $\pm3.2$ \\ 
$B\overline B$ Counting & 
$\pm1.1$ & $\pm1.1$ & $\pm1.1$ & $\pm1.1$ &
$\pm1.1$ & $\pm1.1$ & $\pm1.1$ & $\pm1.1$ \\
Fisher &
$\pm1.5$ & $\pm1.5$ & $\pm1.5$ & $\pm1.5$ &
$\pm1.5$ & $\pm1.5$ & $\pm1.5$ & $\pm1.5$ \\ 
$B\overline B$ likelihood & 
$\pm2.5$ & $\pm2.5$ & $\pm2.5$ & $\pm2.5$ &
$\pm2.5$ & $\pm2.5$ & $\pm2.5$ & $\pm2.5$ \\ 
Model dep.  &
$\pm4.0$ & $\pm4.0$ & $\pm4.0$ & $\pm4.0$ &
$\pm4.0$ & $\pm7.0$ & $\pm4.0$ & $\pm7.0$ \\ 
MC statistics &
$\pm1.1$ & $\pm1.1$ & $\pm1.1$ & $\pm1.1$ &
$\pm1.9$ & $\pm1.6$ & $\pm1.0$ & $\pm1.9$ \\ \hline 
Total &
$\pm6.8$ & $\pm6.6$ & $\pm7.9$ & $\pm7.7$ &
$\pm7.6$ & $\pm9.3$ & $\pm9.5$ & $\pm11.1$ \\ 
\hline\hline
\end{tabular}
\end{center} 
\label{tab:systematic_mult}
\end{table}

The additive systematic errors are uncertainties in the signal
yield from the fit. These arise from three sources:
(1) uncertainties in the signal shapes, (2) uncertainties
in the background shapes, and (3) uncertainties in the
amount of peaking backgrounds. Note that because the 
background shapes and yields float in the fit, much of the 
uncertainty associated with the background is automatically
incorporated in the statistical uncertainty of the signal.
A systematic uncertainty in the background shape is evaluated by
(1) fixing the $m_{\rm ES}$ slopes to the values obtained
from Monte Carlo simulations rather than allowing the slopes to float
in the fit and (2) performing alternative fits in which
the $m_{\rm ES}$ slope is allowed to have a quadratic
dependence on $\Delta E$, as described above. 

The $B^+\to K^+ e^+ e^-$ channel
has one additional systematic uncertainty included. In this
channel we observe two events with the following
property: if the electrons are each assigned pion masses,
then the system is consistent with coming from the decay 
$B^+\to D^0\pi^+$; $D^0\to K^-\pi^+$.
The predicted number of such events, based on measured
probabilities for pions to be misidentified as electrons, is only
about 0.08 event. 
In spite of this small estimated probability we take into account the
possibility that
these events might
in fact be background by including an
asymmetric systematic error corresponding to the
subtraction of two events.
This issue does not arise in the muon channels, where
we have an explicit veto on this background to handle
the much higher probability for a pion to be misidentified
as a muon.

Table~\ref{tab:systematic_additive}
summarizes the additive systematic errors, 
expressed directly in terms of errors on
signal event yields.
Table~\ref{tab:results} gives the total
additive systematic errors propagated to
the branching fractions.
For the combined fits for $B\to K\ell^+\ell^-$ and
$B\to K^*\ell^+\ell^-$ the systematic errors are 
calculated by weighting the systematic errors from
the individual modes based on the expected signal 
from the efficiency and branching fractions.

\begin{table}[h]
 \footnotesize
 \caption[Systematic uncertainties from the fit]{
  The sources of additive systematic uncertainty, expressed directly as 
  uncertainties in the signal yield.
 }
 \begin{center}
 \begin{tabular}{lcccccccc}\hline\hline
  {Systematic} 
  & $K^+e^+e^-$ 
  & $K^+\mu^+\mu^-$ 
  & $K_S^0 e^+e^-$ 
  & $K_S^0 \mu^+\mu^-$ 
  & $K^{*0}e^+e^-$ 
  & $K^{*0}\mu^+\mu^-$ 
  & $K^{*+}e^+e^-$ 
  & $K^{*+}\mu^+\mu^-$ \\
  \hline
Signal shape       &  $\pm0.3$          & $\pm0.0$  &  $\pm0.3$  &  $\pm0.1$  & $\pm0.4$  &  $\pm0.2$  & $\pm0.4$  & $\pm0.3$  \\
Comb bkg. shape    &  $\pm1.8$          & $\pm0.1$  &  $\pm0.7$  &  $\pm0.3$  & $\pm2.6$  &  $\pm0.3$  & $\pm1.3$  & $\pm1.4$  \\
Peaking bkg.       &  $^{+0.0}_{-2.2}$  & $\pm0.7$  &  $^{+0.0}_{-0.1}$ & $\pm0.5$ & $^{+0.2}_{-0.5}$ & $\pm1.2$ & $^{+0.1}_{-0.4}$
& $\pm1.0$ \\
\hline\hline
 \end{tabular}
 \end{center}
 \label{tab:systematic_additive}
\end{table}

\section{Discussion}
\label{sec:discussion}
This analysis has yielded preliminary evidence for the
decay $B\to K\ell^+\ell^-$. The signal events are
primarily in the $B^+\to K^+ e^+ e^-$ channel, although there 
is also an excess in $B^0\to K_S^0 \mu^+\mu^-$.  
The higher event yield in electron over muon channels is not unexpected,
due to the substantially higher detection efficiency for electrons than muons
in the \babar\ detector. 
The significance of this signal, based purely
on the statistical errors, is $5.6\sigma$; when systematic
uncertainties are applied, the estimated significance becomes
$4.4\sigma$. 
We conclude that a signal for $B\to K\ell^+\ell^-$ has been observed.
The measured branching fraction is consistent 
with most predictions, although the central value of 
${\mathcal B}(B\to K\ell^+\ell^-)=(0.78^{+0.24+0.11}_{-0.20-0.18})\times 10^{-6}$
is higher than the Ali~{\it et al.} prediction
of ${\cal B}(B\to K\ell^+\ell^-)=(0.35\pm0.12)\times 10^{-6}$. 
The result is also consistent with Belle's measurement, 
${\cal B}(B\to K\ell^+\ell^-)=(0.75^{+0.25}_{-0.21}\pm0.09)\times 10^{-6}$.
The result from this 77.8 fb$^{-1}$ sample is also higher than the 90\%
C.L.  upper limit obtained from our original analysis of 20.7
fb$^{-1}$; the larger data set includes the smaller data set.
We have made a conservative evaluation of the consistency in our results
as follows: We evaluate
the $B\to K\ell^+\ell^-$ branching fraction using only the data acquired after
that used in the published result. Assuming a true value equal to the result
obtained in this newer dataset, we
find a 3\% probability ($1.9\sigma$) that we would have observed a yield
smaller than our published result from the earlier dataset.

We also observe an excess of events in $B\to K^*\ell^+\ell^-$, with a
significance including systematic uncertainty of
$2.8\sigma$. 
We obtain a central value of
${\mathcal B}(B\to K^*\ell^+\ell^-)=(1.68^{+0.68}_{-0.58}\pm0.28)\times 10^{-6}$
and a 90\% C.L.~upper limit
${\mathcal B}(B\to K^*\ell^+\ell^-)<3.0\times 10^{-6}$. The central
value is within the range expected in the Standard Model, although
the uncertainties are still large.

\section{Conclusions}
\label{sec:Conclusions}

We present preliminary results of our search for $B\to K^{(*)}\ell^+\ell^-$ 
using a sample of $84.4 \times 10^6$
$B\overline B$ pairs produced at the $\Upsilon(4S)$. 
We observe a signal for $B\to K\ell^+\ell^-$ with
\begin{eqnarray}
\BR(B\to K\ell^+\ell^-)&=&(0.78^{+0.24+0.11}_{-0.20-0.18})\times 10^{-6}\nonumber.
\end{eqnarray}
The significance is estimated to be 
$ 4.4\sigma$.
For $B\to K^*\ell^+\ell^-$ we obtain
\begin{eqnarray}
\BR(B\to K^*\ell^+\ell^-)&=&(1.68^{+0.68}_{-0.58}\pm 0.28)\times 10^{-6},\nonumber
\end{eqnarray}
using the constraint $\BR(B\to K^*e^+e^-)/\BR(B\to K^*\mu^+\mu^-)=1.2$.
For the combined $B\to K^*\ell^+\ell^-$ result we quote the 
branching fraction corresponding to the electron channel.
The significance of the \modekstavgll signal is estimated
to be
$2.8\sigma$ 
and we quote a 90\% C.L. limit for this mode of
\begin{eqnarray*}
\BR(B\to K^*\ell^+\ell^-)&<&3.0\times 10^{-6}.
\end{eqnarray*}

\section{Acknowledgments}
\label{sec:Acknowledgments}

We are grateful for the 
extraordinary contributions of our \pep2\ colleagues in
achieving the excellent luminosity and machine conditions
that have made this work possible.
The success of this project also relies critically on the 
expertise and dedication of the computing organizations that 
support \babar.
The collaborating institutions wish to thank 
SLAC for its support and the kind hospitality extended to them. 
This work is supported by the
US Department of Energy
and National Science Foundation, the
Natural Sciences and Engineering Research Council (Canada),
Institute of High Energy Physics (China), the
Commissariat \`a l'Energie Atomique and
Institut National de Physique Nucl\'eaire et de Physique des Particules
(France), the
Bundesministerium f\"ur Bildung und Forschung and
Deutsche Forschungsgemeinschaft
(Germany), the
Istituto Nazionale di Fisica Nucleare (Italy),
the Research Council of Norway, the
Ministry of Science and Technology of the Russian Federation, and the
Particle Physics and Astronomy Research Council (United Kingdom). 
Individuals have received support from 
the A. P. Sloan Foundation, 
the Research Corporation,
and the Alexander von Humboldt Foundation.


\begin{thebibliography}{99}

\bibitem{bib:TheoryA}{A.~Ali {\it et al.}, Phys.~Rev.~D
{\bf 61}, 074024 (2000); P.~Colangelo {\it et al.}, Phys.~Rev.~D
{\bf 53}, 3672 (1996); D.~Melikhov, N.~Nikitin, and S.~Simula,
Phys.~Rev.~D {\bf 57}, 6814 (1998).}

\bibitem{bib:TheoryB}{T.M.~Aliev {\it et al.}, Phys.~Lett.~B {\bf 400},
194 (1997); T.M.~Aliev, M.~Savci, and A.~\"Ozpineci, Phys.~Rev.~D {\bf 56},
4260 (1997); M.~Beneke, Th.~Feldmann, and D.~Seidel, Nucl.~Phys.~B {\bf 612}, 25 (2001);
G.~Burdman, Phys.~Rev.~D {\bf 52}, 6400 (1995); N.G.~Deshpande and J.~Trampetic,
Phys.~Rev.~Lett.~{\bf 60}, 2583 (1988);
C.~Greub, A.~Ioannissian, and D.~Wyler, Phys.~Lett.~B {\bf 346}, 149 (1995);
J.L.~Hewett and J.D.~Wells, Phys.~Rev.~D {\bf 55}, 5549 (1997);
C.Q.~Geng and C.P.~Kao, Phys. Rev. D {\bf 54},
5636 (1996); and references therein.}

\bibitem{bib:AliUpdate}{A.~Ali, E.~Lunghi, C.~Greub, and G.~Hiller, hep-ph/0112300, to appear in Phys.~Rev.~D.}

\bibitem{bib:BaBarKll}{\babar\ Collaboration, B.~Aubert {\it et al.}, Phys.~Rev.~Lett.~{\bf 88}, 241801 (2002).}

\bibitem{bib:Belle}{Belle Collaboration, K.~Abe {\it et al.}, Phys.~Rev.~Lett.~{\bf 88}, 021801 (2002).}

\bibitem{bib:CDF}{CDF Collaboration, T.~Affolder {\it et al.},~\jprl\ {\bf83}, 3378 (1999).}     

\bibitem{bib:CLEO}{CLEO Collaboration, S.~Anderson {\it et al.}, Phys.~Rev.~Lett.~{\bf 87}, 181803 (2001).}     


\bibitem{bib:CLEOKstargam}{CLEO Collaboration, R.~Ammar {\it et al.}, Phys.~Rev.~Lett.~{\bf 71},
674 (1993).}

\bibitem{bib:CLEOXsgam}{CLEO Collaboration, M.S.~Alam {\it et al.}, Phys.~Rev.~Lett.~{\bf 74},
2885 (1995).}

\bibitem{bib:PDG2002}{Particle Data Group, K.~Hagiwara {\it et al.}, Phys.~Rev.~D {\bf 66}, 010001 (2002).}

\bibitem{bib:babarNIM} {\babar\ Collaboration, B.~Aubert {\em et al.}, 
                       Nucl.~Instrum.~Methods A {\bf 479}, 1 (2001).}

\bibitem{bib:Geant4}{{\tt GEANT}4, Geant4 Collaboration, CERN-IT-2002-003,
                     submitted to Nucl.~Instrum.~Methods.}

\bibitem{bib:FoxWolfram}{G.C.~Fox and S.~Wolfram,~Phys.\ Rev.\ Lett.\ {\bf41}, 1581 (1978).} 

\bibitem{bib:Fisher}{R.A.~Fisher, Ann.~Eugenics {\bf 7}, 179 (1936).}

\bibitem{bib:CBShape}{T.~Skwarnicki, ``A Study of the Radiative Cascade
Transitions between the Upsilon-Prime and Upsilon Resonances,'' DESY 
F31-86-02 (thesis, unpublished)(1986).}

\bibitem{bib:ArgusFunction}{ARGUS Collaboration, H.~Albrecht {\it et al.}, Phys.~Lett.~B {\bf 241}, 278 (1990).}

\end{thebibliography}
\end{document}